\documentclass[12pt]{article}
\usepackage{amssymb,amsmath,epsfig,color}
\usepackage[colorlinks=true, pdfstartview=FitV, linkcolor=blue, citecolor=red, urlcolor=magenta]{hyperref}
\allowdisplaybreaks
\begin{document}
\title{\textbf{Stable Configurations of Dark Energy Stars in
$f(R,L_m,T)$ Gravity\thanks{This work is dedicated to Professor
Asghar Qadir on his 80th birthday.}}}
\author{M. Sharif$^{1,2}$\thanks{msharif.math@pu.edu.pk (Corresponding author)} ,
Tayyab Naseer$^{1,3}$\thanks{tayyabnaseer48@yahoo.com (Corresponding
author)}~ and Aqsa Bashir$^1$\thanks{aqsabashir8179@gmail.com}\\
$^1$Department of Mathematics and Statistics, The University of Lahore,\\
1-KM Defence Road Lahore-54000, Pakistan.\\
$^2$Jadara Research Center, Jadara University, Irbid 21110,
Jordan.\\
$^3$Research Center of Astrophysics and Cosmology, Khazar University, \\
Baku, AZ1096, 41 Mehseti Street, Azerbaijan.}

\date{}
\maketitle

\begin{abstract}
In the context of $f(R,L_m,T)$ gravity, two analytical solutions are
proposed in this study, representing the structure of compact
spherical objects whose behavior is dominated by dark energy fluid
parameters. The extended field equations are derived using a static
interior spacetime under the considered model which characterize a
geometric structure including both dark energy and ordinary matter.
The field equations are solved using the connections between dark
energy with conventional matter, which are governed by different
equations of state. This analysis further employs the
Kohler-Chao-Tikekar and Durgapal-Fluoria spacetimes as two distinct
ansatzes to close the system of field equations, resulting in two
independent solutions. For both solutions, the constants are fixed
by matching the interior and exterior spacetimes at the boundary
using junction conditions. This study also assesses the proposed
models using data from SAX J1808.4-3658, and explore the
implications of varying the parametric values within the standard
modified framework. The analysis concludes that both models fulfil
the necessary physical conditions to ensure their candidacy for
realistic stellar structures.
\end{abstract}

\section{Introduction}

A fundamental prediction from Einstein's general relativity (GR),
which provides a thorough mathematical explanation for such objects,
is the existence of ultra-dense celestial bodies. The link between
spacetime curvature, which generates gravity, and the matter
distribution that causes  this curvature is explained by this
theory. According to this theory, the matter-energy content of
spacetime determines its geometry, and the gravitational field that
results is an expression of that geometry. The intrinsic behavior of
celestial objects bound together by their own gravity is defined by
Einstein's non-linear field equations. Consequently, the application
of this modified gravitational framework to the description of
compact astrophysical systems is supported by observational data.
Schwarzschild's innovative solutions, which were initially
calculated in $1916$, indicated an underlying structure of
considerable astrophysical complexity. In order to model a constant
matter-energy density, he suggested two different methods: one
defining the outward geometry \cite{s1} and the other describing the
internal structure \cite{s2}. Further investigation showed that this
solution continued to produce a singularity in the center of the
object. The cosmos is saturated by a powerful repulsive force that
resists gravity and propels its fast expansion, according to later
study. Because the underlying nature of this fascinating force is
still unknown, the scientific community has given  it the name "dark
energy". The cosmological constant is not entirely explained by
Einstein's theory of GR, despite the fact that it provides a
framework for this issue. Many different theories of gravity have
been put forth as a result.

By substituting an arbitrary function $f(R)$ for the Ricci scalar
$R$ in the Einstein-Hilbert action, one obtains $f(R)$ gravity, the
most basic extension of GR. This idea has been thoroughly examined
in the literature as a possible framework for reconciling the
universe's late-time acceleration and early-time inflation
\cite{s3}-\cite{s6}. Numerous studies employing both analytical and
numerical techniques have verified the stability of several $f(R)$
gravity models \cite{s7}-\cite{s11}. A major advancement was made by
Bertolami et al.  through their foundational work, which introduced
matter-geometry coupling into the $f(R)$ formulation \cite{s14}.
This method produced the constitutive framework for $f(R,L_m)$
gravity by adding both the Ricci scalar $R$ and the matter
Lagrangian density $L_m$ to the gravitational action. Significant
changes to the characteristics of compact objects were discovered by
their examination inside this framework, and these findings were
later investigated further by other researchers \cite{s15,s16}.
Harko et al.'s \cite{s17} $f(R,T)$ gravity, which builds the
gravitational action from a generic function of the Ricci scalar $R$
and the energy-momentum trace $T$, was a major extension of GR. It
is possible to take this explicit connection to the trace $T$ as an
indication of exotic matter fields or imperfect fluids that go
beyond the conventional paradigm. The non-zero divergence of the
energy-momentum tensor (EMT), which physically manifests as a
non-geodesic acceleration, is caused by its dependence on the trace
$T$. This extra force causes test particles to undergo non-geodesic
motion, which supports the theory's many cosmological and
astrophysical implications \cite{15}-\cite{taqsa3}.

The $f(R,T)$ gravity poses considerable theoretical and
observational difficulties, yet providies a possible paradigm for
resolving cosmic problems. The theory faces a particular difficulty
in developing viable models that are consistent with both
local-scale gravitational tests and large-scale cosmological
phenomena. Building on this work, Harko and Haghani \cite{22a}
combined the $f(R,T)$ and $f(R,L_m)$ theories into a single
framework that they called $f(R,L_m,T)$ gravity. In this context,
both minimal and non-minimal models have been studied, yielding
outcomes consistent with observational data. According to the study
by Naseer and Said on the internal composition of charged isotropic
stars \cite{22b}, the minimal model yields more favorable results
than the non-minimal counterpart. Several important research studies
\cite{89a}-\cite{100h} support the theoretical and observational
viability of this framework.

The structure and behavior of a self-gravitating system are entirely
determined by the field equations of any gravitational theory. These
equations have higher-order derivatives of geometric quantities,
which makes their solution mathematically challenging. The
application of numerical integration to these equations usually
necessitates the specification of initial or boundary conditions,
even where analytical solutions are known. To fully describe the
system's dynamics, the field equations must be complemented by
additional conditions based on local physical laws. Recent research
indicates that the structural properties of massive objects,
especially compact stellar configurations, can be altered under a
range of physical conditions \cite{57,58}. The gravitational field
of a static, spherical system is described by three independent
equations involving five unknown functions (two from the metric and
three from the matter components). This underdetermination means
that a unique solution is only possible with additional information,
for instance, by prescribing the metric potentials. One can obtain a
well-behaved solution by restricting the metric coefficients,
thereby reducing the number of unknown functions in the system. To
establish a relationship between the two metric functions, the
Karmarkar condition was employed. The condition in question follows
from a differential equation that emerges when the Riemann tensor
components are combined in a specific manner \cite{64}-\cite{67b}.
Under the assumption of a vanishing Weyl tensor, this situation has
also been studied \cite{68}.

Cosmological evidence indicates that only 5\% of the universe is
made up of visible material, with the remaining 26\% and 69\% being
made up of dark matter and dark energy, respectively \cite{z1}.
Regardless of cosmic expansion, dark energy produces a persistent
repulsive force due to its uniform distribution and constant
density. The fundamental gravitational frameworks established by
Newton and Einstein have been modified in an attempt to explain the
mysterious phenomena of dark matter and dark energy, which together
constitute the predominant portion of the universe's matter-energy
content \cite{z2}. In contrast to ordinary and dark matter, dark
energy usually retains a constant density throughout space and time
and has a repulsive gravitational impact within its volume. The
universe's rate of expansion and the creation of massive formations
like galaxy clusters, which are the outcome of gravitational
instabilities, are proof of the influence of dark energy, even
though its origin is still unknown. Astronomers use telescopic
observations of celestial objects over great distances and at
various points in cosmic history to measure this expansion. Every
experiment carried out to date has consistently confirmed the
predictions of GR \cite{z3}. The overall energy density of the
cosmos is extremely close to the critical density necessary for
spatial flatness, according to high-precision observations of the
CMB radiation. There is a strong evidence for accelerated cosmic
expansion based on current astronomical data. A number of phenomena,
including CMB radiation \cite{z6,z7}, the rotational profiles of
galaxies, and observations of type Ia supernovae \cite{z4,z5},
provide important evidence for this expansion.

The repulsive impact of dark energy's negative pressure opposes the
inward gravitational attraction of baryonic matter. To explain dark
energy, two main models have been put forth. The first is the
cosmological constant, which denotes a constant energy density that
permeates the vacuum of space uniformly. The second deals with
dynamic scalar fields, like quintessence, whose energy density
changes over time and space and is not constant. In cosmological
models, the cosmological constant ($\Lambda$) serves as a
description for the contribution of scalar fields that do not vary
with spatial location. If dark energy is described by the
cosmological constant, $\Lambda$, then its measured energy density
in our local universe must be consistent with values derived from
large-scale cosmological observations \cite{z8,z9}. As the primary
component of the cosmos, dark energy could influence and interact
with normal matter in extreme objects like black holes, neutron
stars, and wormholes \cite{z10}. The fundamental role of dark energy
in cosmology motivates the investigation of its potential
astrophysical effects on smaller, local scales.

In cosmology, dark energy is characterized by the equation of state
$p=\omega\rho$, and its nature is defined by the specific value of
the parameter $\omega$. Kiselev \cite{z11} first came up with a
black hole solution that is now widely accepted in the literature as
a typical black hole. It is now commonly accepted as a basic black
hole framework, particularly in scenarios where the parameter
$\omega$ equals $-\frac{1}{3}$.  Additional research on these
well-established black hole solutions, including quantum property
studies, can be found in \cite{z12}-\cite{z16}. Furthermore, studies
investigate wormholes supported by dark energy by modeling it as a
phantom energy with a parameter $\omega$ being less than $-1$
\cite{z17}.  Hubble Space Telescope observations, when analyzed
using an adapted version of the Kahn-Woltjer method, have revealed
important details about how the Milky Way-Andromeda system expands
\cite{z18}. Researchers used a unified analytical method to examine
the Local Group's total dark matter mass as well as its local dark
energy density. Their results offer compelling evidence for the
cosmological constant, bolstering the contemporary understanding of
its role in driving cosmic acceleration.

It is important to clarify the novelty of the present investigation
in relation to existing literature. Although the Kohler-Chao-Tikekar
and Durgapal-Fuloria spacetimes have been employed individually in
various modified gravity contexts such as $f(R,T)$ gravity
\cite{s17}-\cite{20}, $f(R,G)$ theory \cite{22bx}, and gravitational
decoupling schemes \cite{y1,y3}, and the linear $f(R,L_m,T)$ model
has been applied to compact stars with different metric ansatzes
\cite{22b}, the specific combination of these two particular
spacetimes within this framework for a two-fluid (ordinary matter +
dark energy) configuration has not been systematically explored.
Previous studies in this gravity have typically focused on either
anisotropic fluids without an explicit dark energy component or have
employed different metric potentials altogether. Our work fills this
gap by deriving two complete analytical solutions using both the
Kohler-Chao-Tikekar and Durgapal-Fuloria ansatzes under identical
equations of state for dark energy, which provide a transparent
coupling between dark energy and baryonic matter. This unified
framework allows us to \emph{(i)} compare the structural differences
arising from the choice of spacetime geometry within the same
modified gravity setting, \emph{(ii)} derive explicit forms for all
model parameters through the junction conditions, and \emph{(iii)}
conduct a comprehensive multi-criteria stability analysis including
hydrostatic equilibrium, causality, adiabatic index, and Zeldovich
conditions to establish the viability of dark energy stars in this
theory. The present study thus offers new physical insights into how
dark energy can influence the internal structure and stability of
compact objects within the $f(R,L_m,T)$ paradigm, while providing a
systematic benchmark for future investigations.

The structure and characteristics of relativistic dark energy stars
- theoretical objects made of both ordinary matter and dark energy -
are examined in this study. The established existence of dark energy
does not directly affect star formation, as stars are composed
almost entirely of ordinary matter. The literature shows that
understanding these hybrid systems has advanced quickly and
significantly, making this field of study both innovative and
active. The structure of this work is summarized below. The
theoretical framework is covered in Section \textbf{2}, where the
field equations incorporating dark energy are formulated and the
framework's underlying principles are outlined. We describe two
different solutions to the field equations using two particular
ansatzes in Section \textbf{3}. The interior and outside spacetimes
at the boundary are smoothly matched to determine the values of the
constants. The physical characteristics of the derived models are
analyzed in Section \textbf{4}, which includes a detailed
examination of their equilibrium states and stability using a number
of recognized techniques. Section \textbf{5} concludes by
summarizing the key findings of this investigation and discussing
how modified gravity can enhance our understanding of these hybrid
stars.

\section{Fundamentals of $f(R,L_m,T)$ Theory}

A modification to the standard Einstein-Hilbert action defines the
$f(R,L_m,T)$ gravitational framework, which is represented as
\begin{equation}\label{1}
S_{f(R,L_m,T)}=\int\sqrt{-g}\left\{\frac{f(R,L_m,T)}{16\pi}+L_{m}\right\}d^{4}x.
\end{equation}
The symbol $L_m$ represents the Lagrangian density associated with
the matter distribution. When the least-action principle is applied
to Eq.\eqref{1}, the following field equations result
\begin{equation}\label{1a}
G_{\gamma\mu}=8\pi T_{\gamma\mu}^{(eff)},
\end{equation}
where the Einstein tensor is represented by
$G_{\gamma\mu}=R_{\gamma\mu}-\frac{1}{2}R g_{\gamma\mu}$, and
$T_{\gamma\mu}^{(eff)}$ is an effective EMT. The former quantity
becomes
\begin{equation}\label{1b}
T_{\gamma\mu}^{(eff)}=\frac{1}{f_{R}}T_{\gamma\mu}+T_{\gamma\mu}^{(D)}.
\end{equation}
The standard matter EMT is represented by $T_{\gamma\mu}$ on the
right, and the corrections resulting from the action modification
are represented by $T_{\gamma\mu}^{(D)}$. These corrective terms
have the following form
\begin{align}
\nonumber T _{\gamma\mu}^{(D)}&=\frac{1}{8\pi
f_{R}}\bigg[\frac{1}{2}(2f_{T}+f_{L_m}) T_{\gamma\mu}+\frac{1}{2}
({f}-R f_{R})g_{\gamma\mu}\\\label{2}
&-(g_{\gamma\mu}\Box-\nabla_{\gamma}\nabla_{\mu})f_{R}-\frac{1}{2}(2f_{T}+f_{L_m})L_{m}
g_{\gamma\mu}+2f_{T}g^{\xi\zeta}\frac{\partial^2 L_m}{\partial
g^{\gamma\mu}\partial g^{\xi\zeta}}\bigg].
\end{align}

In GR, the line element serves as the foundational tool for
characterizing the spacetime geometry of massive astrophysical
objects, including black holes and neutron stars. Because of this,
the metric offers a crucial basis for comprehending how matter and
energy behave in self-gravitating systems. The static spherically
symmetric metric has a significant analytical advantage since it
greatly simplifies the field equations. The resulting field
equations are made simpler by this spherically symmetric metric,
which makes them easier to solve analytically. We therefore model
the interior spacetime with the following metric
\begin{equation}\label{3}
ds^2=-e^{\sigma(r)} dt^2+e^{\chi(r)} dr^2+r^2(d\theta^2+\sin^2\theta
d\phi^2),
\end{equation}
the fact that this metric's components only depend on the radial
coordinate $r$ makes its static nature clear.

In dense star settings, where factors like rotation, strong magnetic
fields, and intense nuclear interactions generate notable
anisotropies, the assumption of isotropic pressure is frequently
inaccurate. Therefore, a thorough understanding of their dynamics
depends on creating precise models of their interior geometry.
Because this anisotropy influences stellar stability, structure, and
evolution, its inclusion in theoretical models is essential.
Furthermore, a mounting volume of observational evidence suggests
that anisotropic pressure models offer a more compelling
interpretation of the properties observed in selected stellar
phenomena, notably magnetic fields and pulsars \cite{73}-\cite{ts4}.
Our approach is to first model the system as an isotropic fluid,
then incorporate a dark energy term that generates an effective
anisotropic pressure. This configuration is described by the
following EMT
\begin{equation}\label{3a}
T_{\gamma\mu}=(\bar{\rho}+\bar{p_t})k_{\gamma}k_{\mu}+\bar{p_t}
g_{\gamma\mu}-\big(\bar{p_r}-\bar{p_t}\big)V_\gamma V_\mu,
\end{equation}
here, the effective energy density and pressures are defined as
$\bar{\rho}=\rho+\rho^{de}$, $\bar{p_r}=p+p_r^{de}$, and
$\bar{p_t}=p+p_t^{de}$, where $\rho$ is the standard energy density
and $p$ is the isotropic pressure. Additionally, $k_\gamma$ and
$V_\gamma$ represent the four-velocity and four-vector,
respectively. Also, $\rho^{de},~p_t^{de}$, and $p_r^{de}$ correspond
to the dark energy components of the energy density, radial
pressure, and tangential pressure, respectively. When expressed in a
co-moving reference frame, the metric in Eq.\eqref{3} yields the
following results
\begin{equation}\label{3b}
k^\gamma=\big(e^{-\sigma(r)/2},0,0,0\big), \quad
V^\gamma=\big(0,e^{-\chi(r)/2},0,0\big),
\end{equation}
which are governed by the following equations
$$k^\mu V_{\mu}=0, \quad k^\mu
k_{\mu}=-1, \quad V^\mu V_{\mu}=1.$$

Selecting a specific modified gravity model provides a framework to
address GR's shortcomings in describing cosmic phenomena, a
necessary step for advancing cosmological and astrophysical theory.
As a result, choosing a particular modified gravity model is crucial
in modern astrophysics since it directs the interpretation of
observational data as well as the planning of upcoming research. We
utilize the linear $f(R,L_m,T)$ model for this investigation, which
is represented as
\begin{equation}\label{3c}
f(R,L_m,T)=R+\lambda f_1(R)+ 2\alpha f_2(L_m)+\beta f_3(T)=R+2\alpha
L_m + \beta T,
\end{equation}
here, $\beta$ and $\alpha$ are dimensionless variables whose values
will be varied throughout the following study. The $f(R,L_m,T)$
framework provides a more comprehensive understanding of dark energy
when combined with more well-known paradigms such as quintessence.
Classical dark energy cosmic acceleration is usually explained by
paradigms that propose either a steady energy density or a dynamic
field. The $f(R,L_m,T)$ function combines the Ricci scalar, the
matter Lagrangian, and the trace of the EMT into a single framework,
resulting in a more comprehensive geometry-matter interaction.
Precisely explaining the dynamics of compact stars' interiors
requires the application of equations for nonlinear gravitational
fields. These equations provide the fundamental foundation required
to comprehend the structure and evolution of stars when anisotropic
pressure distribution and variations in energy density are taken
into consideration.

In contrast to standard models, which frequently assume specific
forms for the dark energy equation of state, $f(R,L_m,T)$ theory
achieves greater flexibility in the simulation of dark energy
dynamics by accounting for dependencies on both local matter
distribution and geometry. As a result, theoretical predictions made
using this method exhibit better agreement with actual observations
of compact objects. A more reliable evaluation of compact star
stability and sustainability in various astrophysical settings is
made possible by the addition of gravitational potential factors and
anisotropic stress tensors to the updated field equations. Thus, the
$f(R,L_m,T)$ paradigm offers a deeper theoretical foundation for
understanding dark energy than conventional approaches. Equations
\eqref{1a}, \eqref{3a}, and \eqref{3c} provide the field equations
for the system by combining the definitions with the metric from
\eqref{3} as
\begin{align}\nonumber
&e^{-\chi }\left(\frac{\chi
'}{r}-\frac{1}{r^2}\right)+\frac{1}{r^2}=\beta\bigg\{ p-\frac{1}{2}
(p_r^{\text{de}}+p)-(p_t^{\text{de}}+p)+\frac{3}{2}  (\rho
^{\text{de}}+\rho )\bigg\}\\\label{4} &+8 \pi (\rho
^{\text{de}}+\rho )+\alpha  (\rho ^{\text{de}}+\rho ),\\\label{5}
&e^{-\chi } \left(\frac{1}{r^2}+\frac{\sigma
'}{r}\right)-\frac{1}{r^2}=(p_r^{\text{de}}+p) \left(\alpha +\frac{3
\beta }{2}+8 \pi \right)-\frac{\beta}{2}(\rho ^{\text{de}}-2
p_t^{\text{de}}+\rho ),\\\nonumber &\frac{e^{-\chi (r)}}{4} \left(2
\sigma ''-\sigma ' \chi '+\sigma '^2+\frac{2 \sigma '}{r}-\frac{2
\chi '}{r}\right)=\frac{\beta}{2}
(p_r^{\text{de}}+p)-\frac{\beta}{2} (\rho ^{\text{de}}+\rho
)\\\label{6} &+2 \beta (p_t^{\text{de}}+p)+\alpha
(p_t^{\text{de}}+p)+8 \pi (p_t^{\text{de}}+p)-\beta p.
\end{align}
The previously described equations can be used to derive the fluid
parameters in their explicit form.

\section{Formulation of Exact Analytical Models}

Finding solutions to Einstein's field equations in both their
standard and modified forms is a major and continuous difficulty for
physicists. Acknowledging this crucial difficulty requires research
toward workable solutions. Although there are several ways to
address this topic, we concentrate on two specific spacetimes that
are known to perform well in comparable circumstances. Each proposed
solution is examined in detail in the subsequent subsections.

\subsection{Dark Energy Model 1}

The analysis of equations \eqref{4}-\eqref{6}  indicates that the
system is under-determined, as there are seven unknown variables and
an insufficient number of constraining equations. One technique for
solving this system, which has  hardly been discussed in the
literature, treats four of its variables as free parameters. We can
also approach the problem by applying a range of different
constraints simultaneously. A direct proportionality exists between
the density of dark energy and its radial pressure, as defined by
\begin{equation}\label{9}
p_r^{de}=-\rho^{de}.
\end{equation}
Additionally, as the equation demonstrates, the normal density of
baryonic matter is proportional to the density attributed to dark
energy.
\begin{equation}\label{10}
\rho^{de}=\nu\rho,
\end{equation}
with the non-zero constant $\nu$. These two constitutive relations
are not merely phenomenological closures but are deeply rooted in
established theoretical frameworks. The condition \eqref{9}
corresponds to the radial equation of state of the Kiselev dark
energy fluid, which has been extensively studied as the canonical
paradigm for black holes and compact objects embedded in dark energy
environments \cite{z74a}. This represents the cosmological constant
limit ($\omega=-1$) of the dark energy equation of state, which is
the most observationally favored scenario consistent with Type Ia
supernovae, CMB, and BAO data. Furthermore, configurations
satisfying this condition are known as `vacuum-like' and produce de
Sitter-type regularity near the center, preventing singularities
\cite{z74b}. The proportionality \eqref{10}, on the other hand, is
motivated by the interacting dark energy paradigm, which is natural
within the modified framework due to its inherent non-minimal
matter-geometry coupling. This linear relation serves as a
first-order approximation to more complex interacting scenarios and
is consistent with the observational fact that dark energy and
matter energy densities are of comparable magnitude in the present
cosmic epoch. The parameter $\nu$ is subsequently fixed by the
junction conditions at the stellar surface, ensuring consistency
with the exterior Schwarzschild geometry and providing posteriori
validation of these assumptions.

The Kohler-Chao-Tikekar metric offers a fundamental structure for
simulating spherical star systems, an idea that was further
investigated in later astrophysical research \cite{z74,z73}. A
faithful description of pressure anisotropy in compact objects is
necessary for the model to appropriately characterize dense star
systems. Using this metric, Albalahi and his associates \cite{y1}
created spherically symmetric stellar models by converting an
anisotropic seed into an isotropic solution using gravitational
decoupling and the null complexity factor. Siza et al. \cite{y3}
applied the identical ansatz, incorporating a complexity-free
constraint, to formulate models for ultra-relativistic systems.
Within this framework, the metric potentials are defined as follows
\begin{equation}\label{11}
e^{\sigma}=a+b r^2,\quad e^{\chi}=\frac{a+2 b r^2}{a+b r^2},
\end{equation}
for constants $a$ (dimensionless) and $b$ (with dimensions
$\ell^{-2}$). Equations \eqref{4}-\eqref{6} under
\eqref{9}-\eqref{11} yield the explicit expressions for the fluid
variables in their explicit form as
\begin{align}\label{13}
\rho &=\frac{b \left(a (3 \alpha +5 \beta +24 \pi )+2 b r^2 (\alpha
+2 \beta +8 \pi )\right)}{(\alpha +\beta +8 \pi ) (\alpha  \nu
+\alpha +3 \beta  \nu +2 \beta +8 \pi  (\nu +1)) \left(a+2 b
r^2\right)^2},\\\nonumber p&=\big [b \big(a (4 \nu +1) (\alpha +3
\beta +8 \pi )+2 b r^2 (2 \alpha  \nu +\alpha +6 \beta \nu +2 \beta
+8\\\nonumber &\times \pi  (2 \nu +1))\big)\big]\big[(\alpha +\beta
+8 \pi ) (\alpha  \nu +\alpha +3 \beta  \nu +2 \beta +8 \pi  (\nu
+1))\\\label{54} &\times \big(a+2 b
r^2\big)^2\big]^{-1},\\\label{55}\rho^{de}&=\frac{b \nu  \left(a (3
\alpha +5 \beta +24 \pi )+2 b r^2 (\alpha +2 \beta +8 \pi
)\right)}{(\alpha +\beta +8 \pi ) (\alpha  \nu +\alpha +3 \beta  \nu
+2 \beta +8 \pi  (\nu +1)) \left(a+2 b r^2\right)^2},
\\\label{56}p_r^{de}&=-\frac{b \nu  \left(a (3 \alpha +5 \beta +24 \pi )+2 b r^2 (\alpha +2 \beta +8 \pi
)\right)}{(\alpha +\beta +8 \pi ) (\alpha  \nu +\alpha +3 \beta  \nu
+2 \beta +8 \pi  (\nu +1)) \left(a+2 b
r^2\right)^2},\\\label{57}p_t^{de}&=-\frac{b \nu  \left(a (3 \alpha
+5 \beta +24 \pi )+2 b r^2 (\alpha +2 \beta +8 \pi )\right)}{(\alpha
+\beta +8 \pi ) (\alpha  \nu +\alpha +3 \beta  \nu +2 \beta +8 \pi
(\nu +1)) \left(a+2 b r^2\right)^2}.
\end{align}
Wang et al. \cite{nv1} demonstrated that achieving physically or
operationally viable configurations requires detailed mathematical
validation of stability across the entire parameter space. Xiang at
el. \cite{nv2} highlighted how theoretical models and observational
catalogs of high-energy sources are essential for building a
complete picture of the extreme universe. Recent advances in
non-linear systems and multibody dynamics collectively highlight the
importance of robust analytical and numerical frameworks for
stability assessment \cite{nv3}-\cite{nv6}.

\subsubsection{Applications of Boundary Conditions}

The behavior of field equations and their derivatives at the
interfaces separating different materials or regions is defined by
junction conditions, which are essential for constructing realistic
models of astrophysical objects like stars. These conditions, which
include the continuity of both the metric and extrinsic curvature,
provide the consistency of the field equations and a smooth matching
between adjacent spacetime sectors. These conditions are essential
for solving boundary value problems, modeling wave and disturbance
propagation, and describing interfacial interactions in physical
systems. For a complete model, the interior metric of an object is
paired with the Schwarzschild solution as its exterior spacetime
\eqref{3} as
\begin{equation}\label{18}
ds^2=-\bigg(1-\frac{2M}{r}\bigg)dt^2
+\bigg(1-\frac{2M}{r}\bigg)^{-1}dr^2+r^2(d\theta^2+\sin^2\theta
d\phi^2),
\end{equation}
with $r>2M$, where the total mass of the star is represented by $M$.

The fundamental framework for comprehending the key features of
astrophysical processes in spherical geometries is provided by the
boundary conditions \cite{41ja}. Two main categories are used to
classify these conditions. The first category enforces continuity of
the metric tensor, which guarantees that the inherent geometry
extends beyond the boundary. Continuity of the extrinsic curvature
constitutes the second prerequisite, determining the degree of
smoothness with which the two regions are matched. This condition
guarantees that the pressure will disappear at a finite radial
coordinate $r=R$, where $R$ stands for the stellar radius. The
following three equations result from these conditions
\begin{align}\label{18a}
1-\frac{2 M}{R}&=a+b R^2,\\\label{49} 1-\frac{2 M}{R}&=\frac{a+b
R^2}{a+2 b R^2}, \\\label{50} p=0&=a (4 \nu +1) (\alpha +3 \beta +8
\pi )+2 b r^2 (2 \alpha\nu +\alpha +6 \beta \nu +2 \beta +8\pi(2\nu
+1)).
\end{align}
The triplet ($a,b,\nu$) is expressed as follows when these equations
are solved concurrently
\begin{align}\label{19}
a&=\frac{R-4 M}{R},\\\label{52} b&=\frac{2 M}{R^3},
\\\label{53}\nu&=\frac{-4 \beta  M+\alpha  R+3 \beta  R+8 \pi  R}{4
(\alpha +3 \beta +8 \pi ) (2 M-R)}.
\end{align}
Following that, these values will be used for the visual evaluation
of the answer that was derived. A more thorough examination of how
the amended theory affects the physical characteristics of stellar
models will be made easier by incorporating the parameter $\alpha$
into these constants.

\subsection{Dark Energy Model 2}

To close the system \eqref{4}-\eqref{6}, we use a different
particular metric potential form that was first proposed by Durgapal
and Fuloria \cite{z75}. This can be formulated as follows
\begin{equation}\label{22}
e^{\sigma}=y\big(r^2 z+1\big)^4,\quad e^{\chi}=\frac{7 \left(r^2
z+1\right)^2}{7-r^4z^2-10 r^2 z},
\end{equation}
with $y$ (dimensionless) and $z$ (having $\ell^{-2}$ as dimension)
being constants. This solution has been widely utilized by
researchers to investigate various physical characteristics of
astrophysical systems. The model described above served as the
foundation for several works in compact stellar physics under the
minimal geometric deformation approach. Within the framework of
modified gravity, the gravitational decoupling method was employed
by Sharif and Saba \cite{z77} to analyze the isotropic
Durgapal-Fuloria stellar model. Ama-Tul-Mughani et al. \cite{z78}
showed that the extended gravitational decoupling method is
effective for studying the internal composition of compact stars,
which expands the solution space for models like the
Durgapal-Fuloria ansatz.

After replacing \eqref{9} and \eqref{10} under the spacetime
\eqref{22}, Eqs.\eqref{4}-\eqref{6} become
\begin{align}\nonumber
\rho&=\big[4 z \big(2 \alpha \big(r^4 z^2+2 r^2 z+9\big)+\beta
\big(r^4 z^2-8 r^2 z+31\big)+16 \pi \big(r^4 z^2+9\\\nonumber &+2
r^2 z\big)\big)\big]\big[7 (\alpha +\beta +8 \pi ) \big(r^2
z+1\big)^3 (\alpha(1+  \nu) +\beta(  3\nu +2) +8 \pi (\nu
\\\label{26} &+1))\big]^{-1},\\\nonumber p&=-\big[4 z \big(2 \alpha  \big((\nu +2) r^4 z^2-13
\nu +2 (6 \nu +7) r^2 z-4\big)+\beta \big(-3 (26 \nu\\\nonumber &
+7)+(6 \nu +5) r^4 z^2+8 (9 \nu +5) r^2 z\big)+16 \pi \big(-13 \nu
+(\nu +2) r^4 z^2\\\nonumber &+2 (6 \nu +7) r^2
z-4\big)\big)\big]\big[7 (\alpha +\beta +8 \pi ) \left(r^2
z+1\right)^3 (\alpha  \nu +\alpha +3 \beta \nu\\\label{27} & +2
\beta +8 \pi (\nu +1))\big]^{-1},\\\nonumber \rho^{de}&=\big[4 \nu z
\big(2 \alpha  \big(r^4 z^2+2 r^2 z+9\big)+\beta  \big(r^4 z^2-8 r^2
z+31\big)+16 \pi  \big(r^4 z^2\\\nonumber &+2 r^2
z+9\big)\big)\big]\big[7 (\alpha +\beta +8 \pi )\big(r^2 z+1\big)^3
(\alpha  \nu +\alpha +3 \beta  \nu +2 \beta +8\\\label{28} &\times
\pi (\nu +1))\big]^{-1} ,\\\nonumber p_r^{de}&=-\big[4 \nu z \big(2
\alpha \big(r^4 z^2+2 r^2 z+9\big)+\beta  \big(r^4 z^2-8 r^2
z+31\big)+16 \pi  \big(r^4 z^2\\\nonumber &+2 r^2
z+9\big)\big)\big]\big[7 (\alpha +\beta +8 \pi )\big(r^2 z+1\big)^3
(\alpha  \nu +\alpha +3 \beta  \nu +2 \beta +8\\\label{29} &\times
\pi (\nu +1))\big]^{-1},\\\nonumber p_t^{de}&=-\big[4 \nu z \big(2
\alpha \big(r^4 z^2+2 r^2 z+9\big)+\beta  \big(r^4 z^2-8 r^2
z+31\big)+16 \pi  \big(r^4 z^2\\\nonumber &+2 r^2
z+9\big)\big)\big]\big[7 (\alpha +\beta +8 \pi )\big(r^2 z+1\big)^3
(\alpha  \nu +\alpha +3 \beta  \nu +2 \beta +8\\\label{30} &\times
\pi (\nu +1))\big]^{-1}.
\end{align}

\subsubsection{Junction Conditions}

The spacetime must be static and asymptotically flat in order to
meet the matching criteria for the exterior of a spherically
symmetric star, whose metric functions only depend on the radial
coordinate. For the first solution, our analysis employed the
Schwarzschild metric given in Eq.\eqref{18}. At the boundary $r=R$,
the two fundamental conditions produce the following set of
equations
\begin{align}\label{32}
1-\frac{2 M}{R}&=y \left(R^2 z+1\right)^4,\\\label{32b} 1-\frac{2
M}{R}&=\frac{7-R^4z^2-10 R^2 z}{7 \left(R^2 z+1\right)^2},
\\\nonumber p=0&=4 z \big(2 \alpha  \big((\nu +2) r^4 z^2-13
\nu +2 (6 \nu +7) r^2 z-4\big)+\beta \big(-3 (26 \\\nonumber &\times
\nu +7)+(6 \nu +5) r^4 z^2+8 (9 \nu +5) r^2 z\big)+16 \pi \big(-13
\nu +(\nu \\\label{32a} &+2) r^4 z^2+2 (6 \nu +7) r^2 z-4\big)\big).
\end{align}
Equations \eqref{32}-\eqref{32a} simultaneously yield the three
constants as
\begin{align}\label{33}
y&=\frac{R-2 M}{R \left(R^2 z+1\right)^4},\\\label{34}z &= \frac{-7
M R^2-2 \sqrt{9 R^6-14 M R^5}+6 R^3}{7 M R^4-4 R^5},\\\nonumber \nu
&= \big[8 \alpha +21 \beta -4 \alpha  R^4 z^2-5 \beta  R^4 z^2-32
\pi R^4 z^2-28 \alpha  R^2 z-224 \pi  R^2 z\\\label{35} &-40 \beta
R^2 z+64 \pi\big]\big[2 (\alpha +3 \beta +8 \pi ) \big(R^4 z^2+12
R^2 z-13\big)\big]^{-1}.
\end{align}

\section{Physical Analysis of the Resulting Solutions}

This section provides a thorough evaluation of the suggested models,
highlighting their key features and stability. This study aims to
clarify the stability of the solutions by looking at how they react
to different outside factors. We first examine basic physical
properties, such as the constancy of metric coefficients and matter
variables, in order to assess the viability of the solutions in
realistic astrophysical contexts. The next subsections will give a
comprehensive description of the methods and tests used to evaluate
these attributes. Five distinct magnitudes of the parameter $\alpha$
such as ($0, 0.4, 0.8, 1.2, 1.6$) together with two choices of
$\beta$ ($1.3$ and $1.9$) are applied in the graphical analysis. The
physical properties under investigation belong to the stellar object
SAX J1808.4-3658 \cite{89fa}, characterized by a mass of $M = 0.9
\pm 0.09 M_{\bigodot}$ and a radius of $R = 7.951 \pm 0.2\text{
km}$. We evaluate how the creation of stellar bodies is influenced
by modified $f(R,L_m,T)$ gravity, as illustrated in the graphical
interpretation that follows.

\subsection{Metric Functions and Matter Determinants}

The metric potentials must be contained inside the interval $(0,1]$
at the origin and be monotonically increasing functions from the
star's center outward, $r=0$. Additionally, both functions must be
positive and finite throughout the whole domain that is being
examined. The graphical behavior of these potentials for both models
is shown in Figure \textbf{1}, demonstrating their regularity across
the domain and non-singularity at the center.
\begin{figure}[h!]\center
\epsfig{file=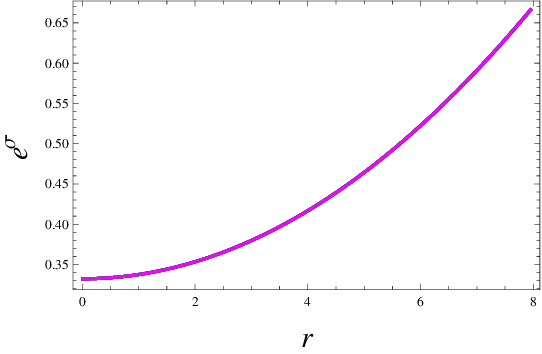,width=0.45\linewidth}\epsfig{file=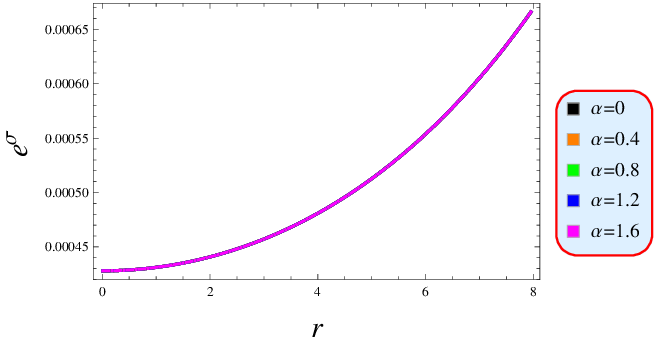,width=0.55\linewidth}
\epsfig{file=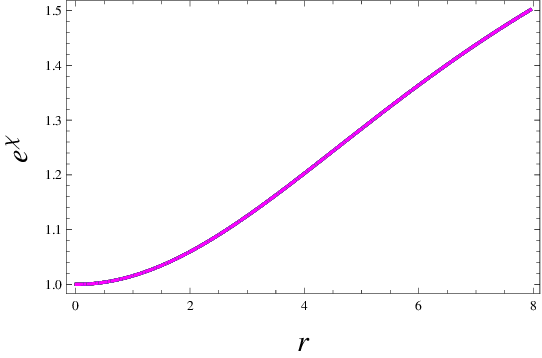,width=0.45\linewidth}\epsfig{file=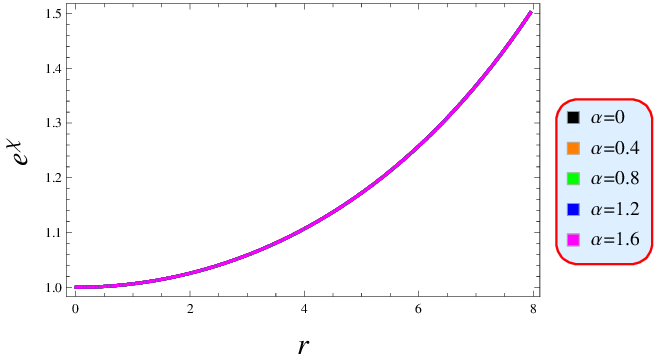,width=0.55\linewidth}
\caption{Geometric elements \eqref{11} and \eqref{22} for solutions
1 and 2 corresponding to left and right plots, respectively under
$\beta=1.3$ (solid) and $\beta=1.9$ (dashed).}
\end{figure}

Analyzing the fundamental physical parameters, such as energy
density and pressure, is necessary before looking at the structure
and stability of the matter distribution. Three requirements must be
met by the energy density and pressure for a star to be physically
viable: they must be positive, finite everywhere, and show an
outward trend that decreases monotonically. Figure \textbf{2} shows
the profiles for both quantities, demonstrating consistent and
physically acceptable behavior. Model 1 exhibits greater density
than Model 2, indicating that the first configuration yields a more
compact object. Furthermore, both the pressure and energy density
are found to decrease as the parameter $\alpha$ increases. The
pressure is seen in this figure to vanish at the stellar surface
$r=R$, which agrees with the anticipated boundary behavior. As one
gets farther from the center, the dark energy density monotonically
drops and is always positive. That dark energy generates a repulsive
force driving cosmic acceleration is reinforced by the behavior of
its radial pressure: it is negative throughout and increases
monotonically. The tangential pressure associated with dark energy
maintains positivity while decreasing smoothly from the central
region. Figure \textbf{3} presents these profiles, confirming that
the calculated behavior matches theoretical expectations.
\begin{figure}[h!]\center
\epsfig{file=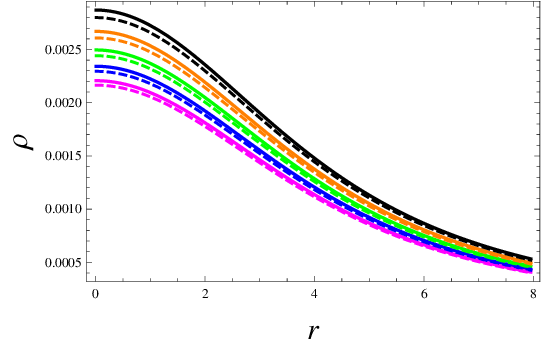,width=0.45\linewidth}\epsfig{file=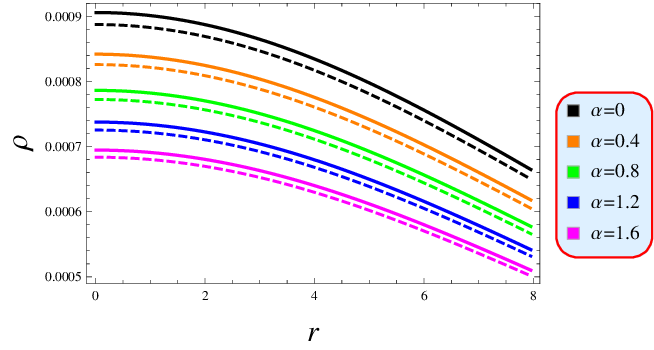,width=0.55\linewidth}
\epsfig{file=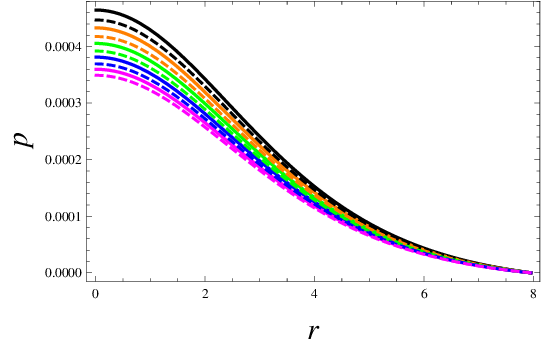,width=0.45\linewidth}\epsfig{file=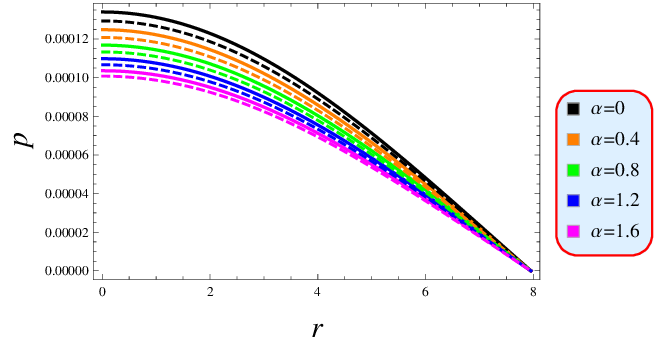,width=0.55\linewidth}
\caption{Matter variables for solutions 1 and 2 corresponding to
left and right plots, respectively under $\beta=1.3$ (solid) and
$\beta=1.9$ (dashed).}
\end{figure}
\begin{figure}[h!]\center
\epsfig{file=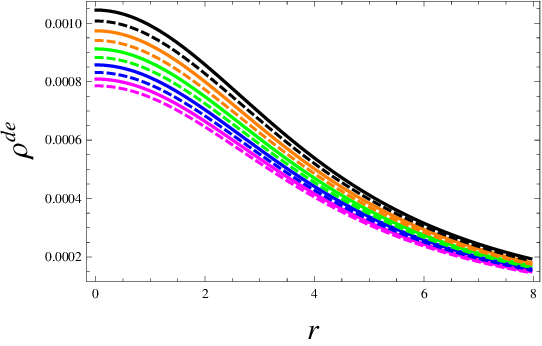,width=0.45\linewidth}\epsfig{file=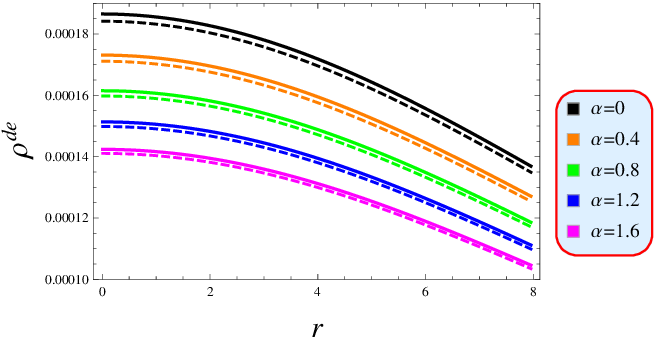,width=0.55\linewidth}
\epsfig{file=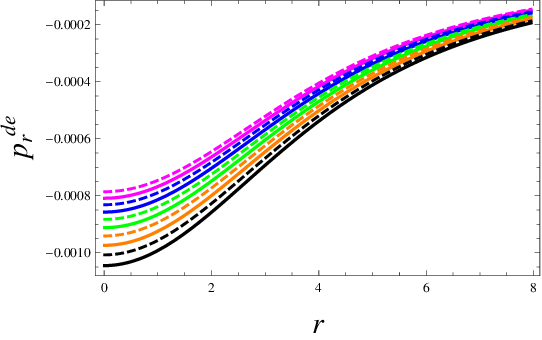,width=0.45\linewidth}\epsfig{file=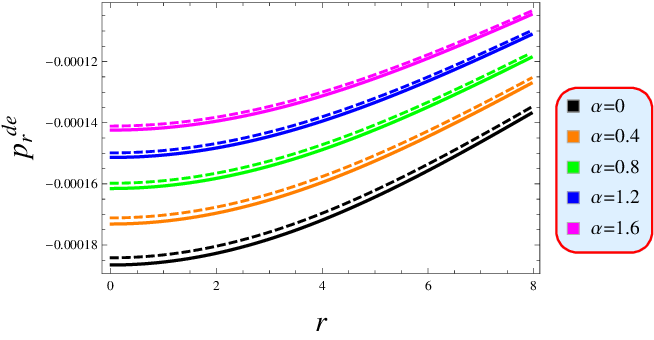,width=0.55\linewidth}
\epsfig{file=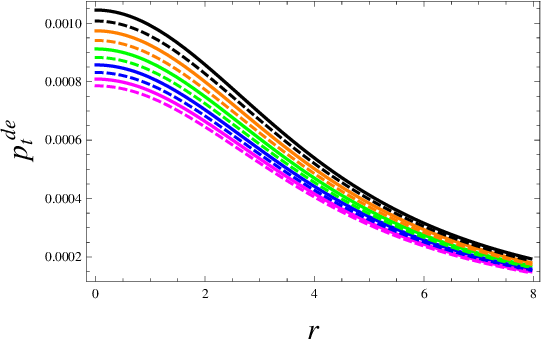,width=0.45\linewidth}\epsfig{file=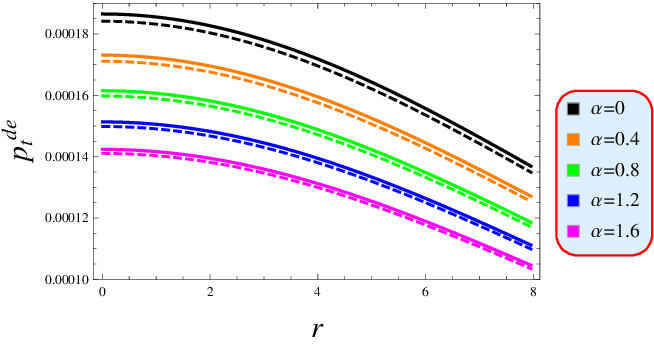,width=0.55\linewidth}
\caption{Dark energy matter variables for solutions 1 and 2
corresponding to left and right plots, respectively under
$\beta=1.3$ (solid) and $\beta=1.9$ (dashed).}
\end{figure}
\begin{figure}[h!]\center
\epsfig{file=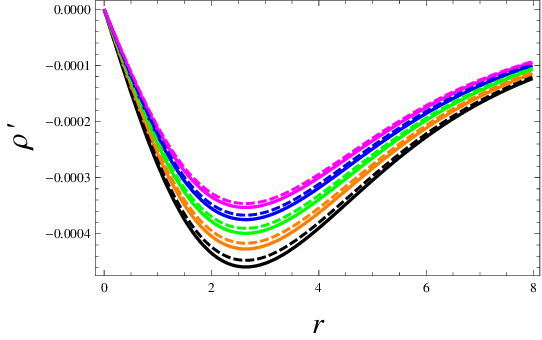,width=0.45\linewidth}\epsfig{file=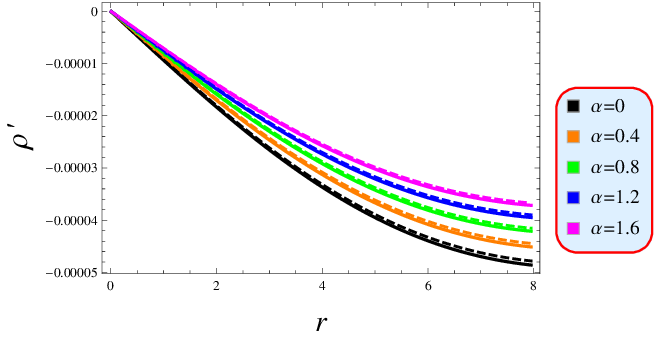,width=0.55\linewidth}
\epsfig{file=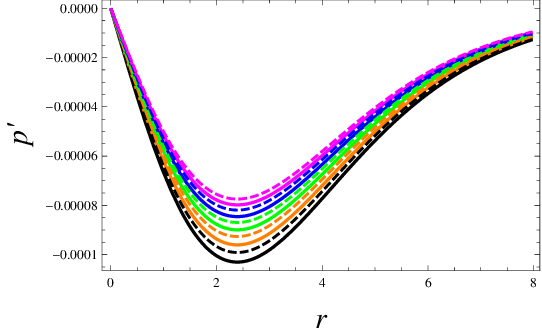,width=0.45\linewidth}\epsfig{file=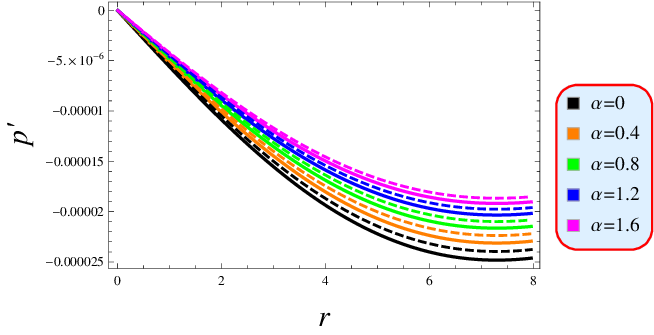,width=0.55\linewidth}
\caption{Gradient for solutions 1 and 2 corresponding to left and
right plots, respectively under $\beta=1.3$ (solid) and $\beta=1.9$
(dashed).}
\end{figure}

Evaluating the feasibility of the solution necessitates an
examination of the radial gradients, which indicates the rate of
change of various physical characteristics. The boundary condition
$\rho'_{r=0}=0=p'_{r=0}$ represents this. Figure \textbf{4}
illustrates how the pressure and energy density gradients behave.
The figure demonstrates that both solutions show consistent trends
and meet the necessary requirements.

\subsection{Mass and Some Other Physical Parameters}

Using the relation, we calculate the mass contained by a spherical
surface of radius $r$
\begin{equation}
m(r)=4\pi\int_0^r r^2\rho(r)dr.
\end{equation}
A differential equation can be solved analytically or numerically,
depending on how complex the energy density profile is. As $r$ gets
closer to the origin, the mass function $m(r)$ converges to zero, as
seen in Figure \textbf{5}. Additionally, in the outer radial
direction, the function shows a strictly monotonically growing
behavior.

An object's compactness, which is based on the mass to radius ratio,
is a crucial factor in controlling its gravitational binding energy
and overall stability. Because of their massive gravitational
fields, extremely compact objects like neutron stars drastically
alter the surrounding spacetime. The compactness parameter of a
stable compact object, stated to be
$\big(\mu(r)=\frac{m(r)}{r}\big)$, must not be greater than $0.44$
for stability, according to Buchdahl's \cite{42} theorem. This bound
was initially developed for spherically symmetric, isotropic
systems, but subsequent studies have expanded its use to include
self-gravitating anisotropic structures. When a photon escapes a
massive object's gravitational field, it loses energy and
experiences surface redshift, which causes its wavelength to
noticeably rise. This is characterized by
\begin{equation}
z_s=\big\{1-2\mu(r)\big\}^{-\frac{1}{2}}-1.
\end{equation}
The internal redshift shows a positive correlation with compactness,
indicating a joint increase in both quantities. The highest
permitted surface redshift is increased when anisotropy is present.
For isotropic matter, this limit is less than 2 \cite{42}, but for
anisotropic distributions, it rises to $5.211$ \cite{89g}. To create
a workable solution, this restriction must be met. Figure \textbf{5}
shows that the profiles for both parameters remain physically
feasible throughout the domain.
\begin{figure}[h!]\center
\epsfig{file=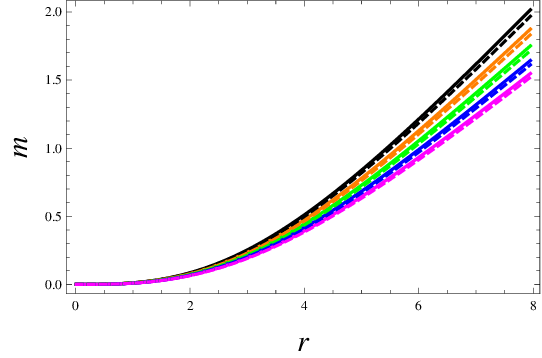,width=0.45\linewidth}\epsfig{file=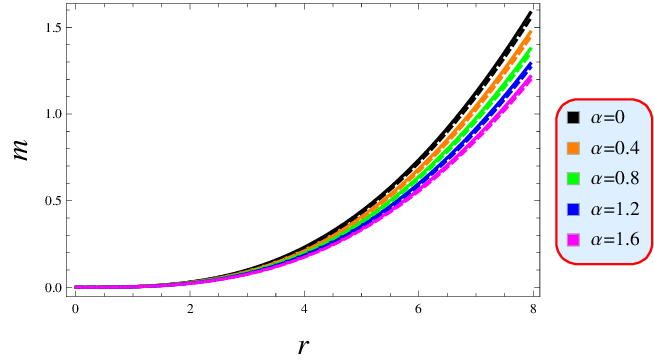,width=0.55\linewidth}
\epsfig{file=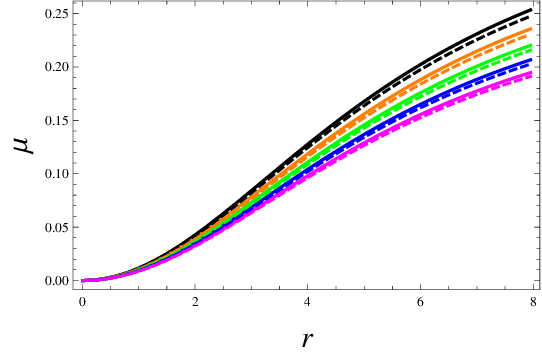,width=0.45\linewidth}\epsfig{file=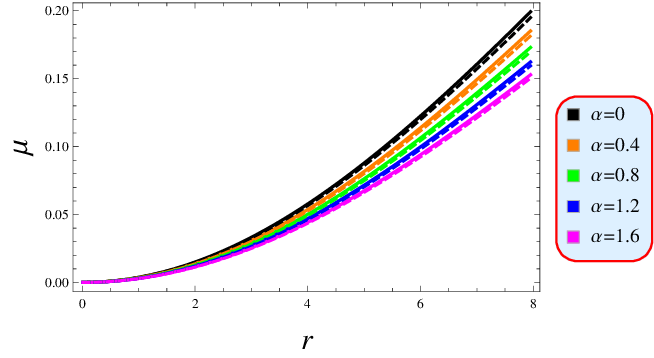,width=0.55\linewidth}
\epsfig{file=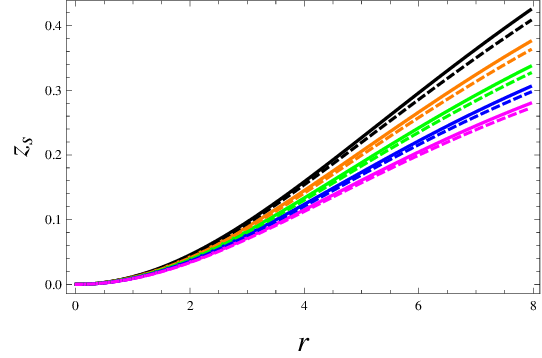,width=0.45\linewidth}\epsfig{file=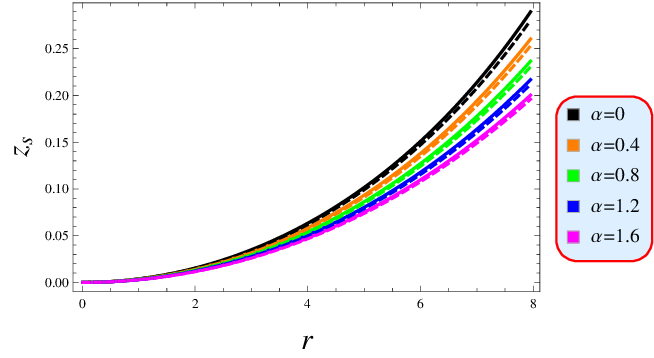,width=0.55\linewidth}
\caption{Mass, compactness and surface redshift for solutions 1 and
2 corresponding to left and right plots, respectively under
$\beta=1.3$ (solid) and $\beta=1.9$ (dashed).}
\end{figure}

\subsection{Viability and Equation of State}

If a star's interior satisfies certain characteristics, it will be
viable. These circumstances are known as the energy conditions, and
they can be broadly classified as dominant, weak, strong, or null.
They are used to identify whether a solution's matter content is
exotic or physically common. By applying these restrictions, the
permitted pressure and energy density values are established,
ensuring consistency with empirical observations. These constraints
are mathematically expressed as
$$\rho\geq0, \quad \rho+p\geq0, \quad \rho-p\geq0, \quad \rho+3p\geq0.$$
We must confirm that our models meet these requirements in order to
determine their physical acceptability. The model's physical
acceptability is ensured by the consistently positive profiles shown
in Figure \textbf{6} for all pertinent parameters, which verify that
the star interior is made up of normal fluid.
\begin{figure}[h!]\center
\epsfig{file=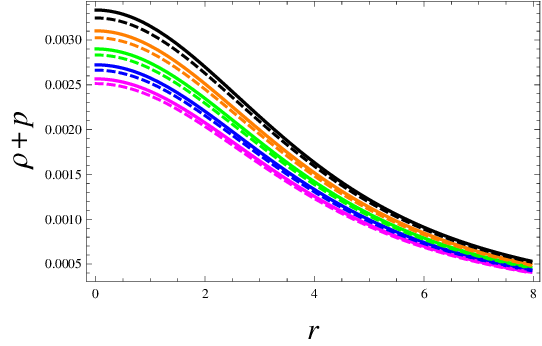,width=0.45\linewidth}\epsfig{file=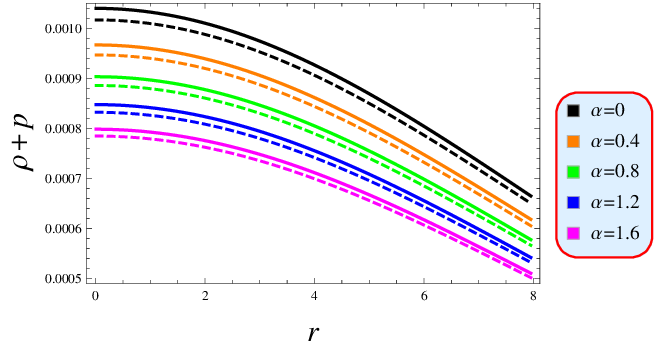,width=0.55\linewidth}
\epsfig{file=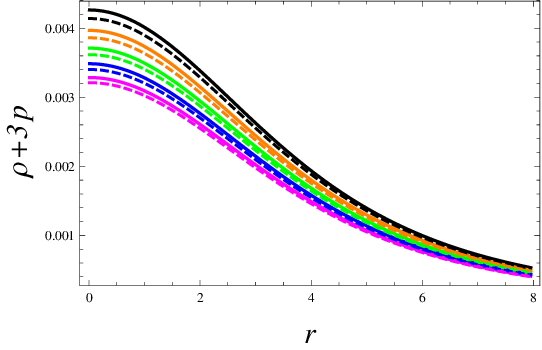,width=0.45\linewidth}\epsfig{file=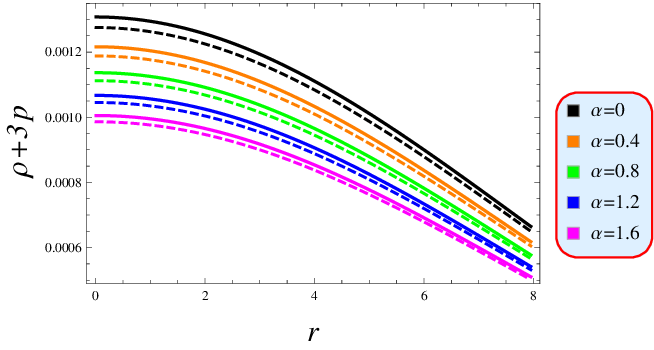,width=0.55\linewidth}
\epsfig{file=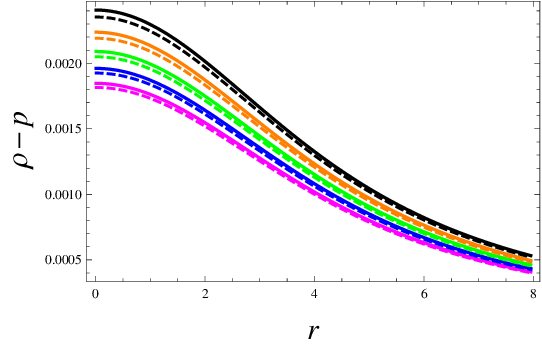,width=0.45\linewidth}\epsfig{file=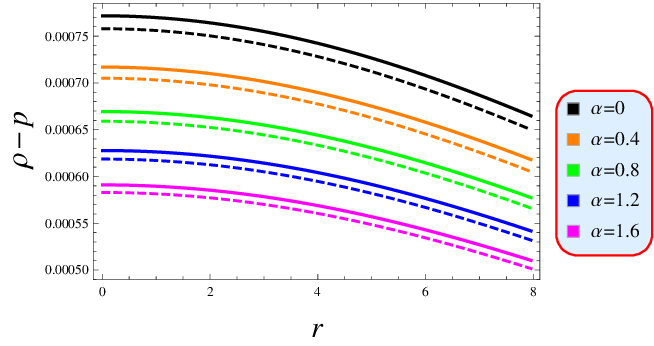,width=0.55\linewidth}
\epsfig{file=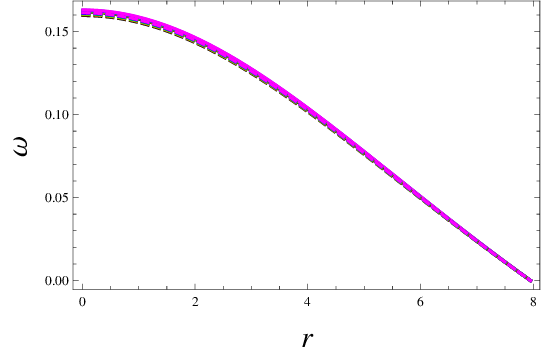,width=0.45\linewidth}\epsfig{file=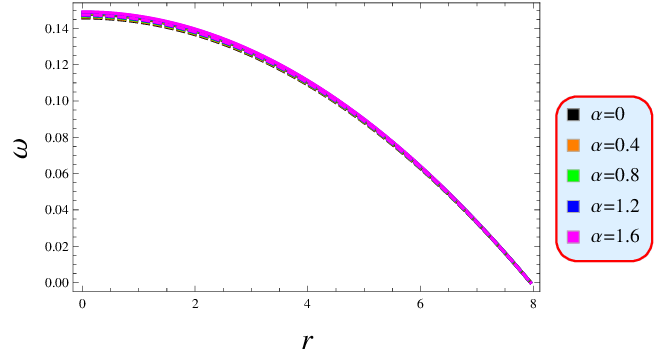,width=0.55\linewidth}
\caption{Energy conditions and realistic parameter for solutions 1
and 2 corresponding to left and right plots, respectively under
$\beta=1.3$ (solid) and $\beta=1.9$ (dashed).}
\end{figure}

According to the formula $p=\omega\rho$, the pressure $p$ in a
barotropic fluid is exactly proportional to the energy density
$\rho$. A fluid must have a pressure to energy density ratio that is
bounded within the interval $[0,1]$ to be physically viable.
Consistency with the basic thermodynamic principles regulating star
interiors is ensured by meeting this requirement. The last plot in
Figure \textbf{6} shows that the parameter $\omega$ stays below
unity across the domain. This pattern shows that density
predominates over pressure in the self-gravitating system, causing
all fluid characteristics to systematically decrease outward toward
the stellar boundary.

\subsection{Observational Validation Through Mass-Radius Diagram}

The mass-radius relation is crucial in understanding the structure,
composition, and stability of compact stars. Chandrasekhar's
ground-breaking study \cite{58aa,59aa} established the maximum mass
for white dwarfs. In a 1931 research paper, he used relativistic
electron degeneracy pressure to estimate the mass limit of white
dwarfs at around $1.4 M_{\bigodot}$ \cite{58aa}. In his second work
\cite{59aa}, Chandrasekhar addressed the consequences of massive
star collapse into neutron stars and black holes, building on this
fundamental revelation. These pioneering papers established the
theoretical foundation for comprehending compact stellar remnants.
Despite the Chandrasekhar limit's robustness, some observational
data call its universality into question. The supernova SNLS-03D3bb
exhibited features of a progenitor with a mass greater than
Chandrasekhar, indicating a super Chandrasekhar white dwarf
\cite{62aa}. Massive neutron stars, like the one having mass $1.97
M_{\bigodot}$ observed via Shapiro delay \cite{4y} and the other
with the mass $2.01 M_{\bigodot}$ in a compact binary system
\cite{64aa}, present challenges to standard equations of state.
Observations of white dwarfs up to $2.4 M_{\bigodot}$ \cite{65aa}
and neutron stars up to $2.14 M_{\bigodot}$ \cite{66aa} suggest that
stellar mass constraints can vary depending on physical conditions.

Several theoretical models have been presented to account for these
anomalous observations. The TOV equations are used to model static
fluid spheres in GR \cite{z80}, while constrained neutron star
masses are studied using various equations of state \cite{68aa}. The
Chandrasekhar mass in other gravity theories might vary based on
model factors \cite{79aa}. Gravitational wave observations helped to
determine the maximum mass of neutron stars \cite{80aa}. Here, we
use our newly developed theoretical models, which are based on
observational data from three different stellar candidates to
examine the link between the mass and radius of compact stars. In
the context of GR, this study is crucial for confirming the
precision of our models. The mass-radius relation for particular
parameters is shown in Figure \textbf{7}, which allows us to
accurately estimate the radii of the suggested compact star
candidates. To further clarify the behavior of these compact
objects, we also present a thorough overview of the numerical
results for different compact stars. For both solutions, three
different stars have the following predicted radii provided in Table
\textbf{1}.
\begin{table}[h!] \scriptsize \centering
\caption{Predicted radii (in $km$) of different pulsars for specific
values of $\alpha$ and $\beta$.}\label{t2} \vspace{+0.07in}
\setlength{\tabcolsep}{0.95em} \begin{tabular}{cccccc}
\hline\hline  & \textbf{LMC~X-4} & \textbf{XTE~J1814-334} &
\textbf{SMC~X-4}
\\\hline
\textbf{Solution 1 ($\alpha=0.4$ and $\beta=1.3$)} & 8.91 & 8.89 &
8.86
\\\hline
\textbf{Solution 1 ($\alpha=0.8$ and $\beta=1.3$)} & 8.83 & 8.81 &
8.78
\\\hline
\textbf{Solution 2 ($\alpha=0.4$ and $\beta=1.3$)} & 8.74 & 8.73 &
8.76
\\
\hline\hline
\end{tabular}
\end{table}
\begin{figure}[h!]\center
\epsfig{file=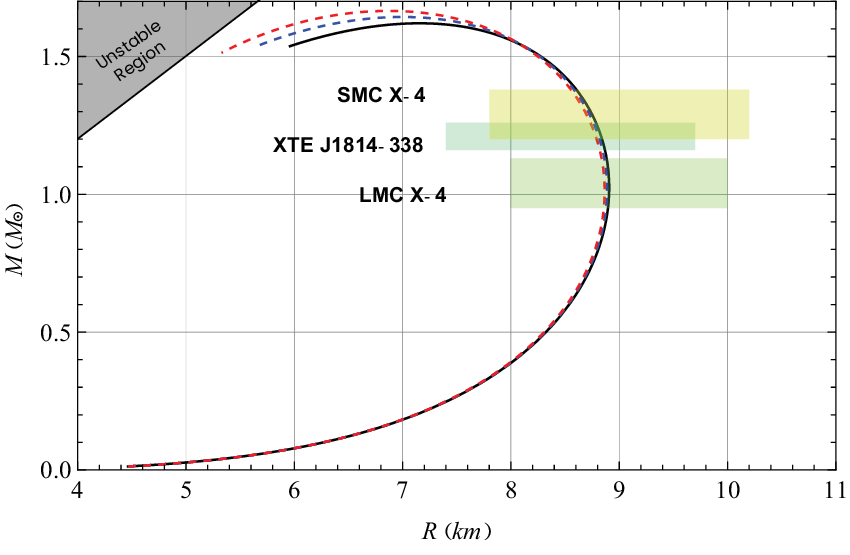,width=0.55\linewidth} \caption{Mass-radius
diagram for solution 1 with $\alpha=0.4$ (red) \& $\alpha=0.8$
(blue), and for solution 2 with $\alpha=0.4$ (black) under
$\beta=1.3$.}\label{f10}
\end{figure}

\subsection{Stability Analysis}

A key component of their research is determining whether a compact
star model is physically feasible by looking at how stable its
solutions are. In order to comprehend how compact stars behave in
the presence of perturbations, the following research focuses on
their equilibrium and stability. Scientists can determine the
parameter space controlling a star's stability and transition to
dynamical evolution by using a variety of mathematical techniques.
Several of these tests are used in the analysis that follows.

\subsubsection{Hydrostatic Equilibrium}

The equilibrium state of the suggested models is evaluated in the
analysis that follows by taking into account the contributions of
all pertinent forces. A star in hydrostatic equilibrium is stable
because the expansion force of internal pressure and the compressive
force of gravity are balanced. Quantum phenomena like electron or
neutron degeneracy pressure maintain this equilibrium condition in
compact star objects. In the framework of general relativity, the
Tolman-Oppenheimer-Volkoff equations are essential for examining
hydrostatic equilibrium \cite{z80,z81}. This equilibrium maintains
internal pressure to oppose gravity without violating the laws of
physics. When discussing $f(R,L_m,T)$ gravity, equations are stated
as
\begin{align}\nonumber
\nabla^\gamma{T}_{\gamma\mu}&=\frac{1}{8 \pi +
2f_{T}+f_{L_m}}\bigg[\nabla_\mu\{(2f_{T}+f_{L_m})L_{m}\}-{T}_{\gamma\mu}\nabla^\gamma
(2f_{T}+f_{L_m})\\\label{64} &-(f_{T}\nabla_\mu T+f_{L_m}\nabla_\mu
L_m)-8\pi\nabla^\gamma\xi_{\gamma\mu}-4g^{\zeta\nu}\nabla^\gamma\bigg(f_{T}\frac{\partial^2
L_{m}}{\partial g^{\gamma\mu}\partial g^{\zeta\nu}}\bigg)\bigg] .
\end{align}
This suggests the presence of an additional force that disrupts the
geodesic track of particles. Using the model \eqref{3c}, this
equation is transformed into the following for the intrinsic
geometry \eqref{3}, given as
\begin{align}
\frac{dp}{dr}+\frac{1}{2}
\sigma'(\rho+p)-\frac{2(p_t^{de}-p_r^{de})}{r}+\frac{dp_r^{de}}{dr}+\frac{\alpha}
{2\alpha+2\beta+8\pi}\bigg(\frac{d\bar{p}_r}{dr}+2\frac{d\bar{p}_t}{dr}+\frac{d\bar{\rho}}{dr}\bigg)=0.
\end{align}
This may be compressed into a more compressed form as
\begin{align}
f_{total}\equiv f_g +f_h +f_d +f_m =0,
\end{align}
with \begin{eqnarray*} f_g&=&-\frac{1}{2}\sigma'(\rho+p),\quad
f_d=\frac{2(p_t^{de}-p_r^{de})}{r}-\frac{dp_r^{de}}{dr},\\
f_h&=&-\frac{dp}{dr}, \quad f_m=-\frac{\alpha}
{2\alpha+2\beta+8\pi}\bigg(\frac{d\bar{p}_r}{dr}+2\frac{d\bar{p}_t}{dr}+\frac{d\bar{\rho}}{dr}\bigg).
\end{eqnarray*}
The equilibrium requirement is satisfied throughout the radial
region, as shown by the displayed forces in Figure \textbf{8}. This
outcome confirms that the suggested models consistently maintain
hydrostatic equilibrium. Thus, the application of this modified
gravitational framework to the description of compact astrophysical
systems is supported by actual data.
\begin{figure}[h!]\center
\epsfig{file=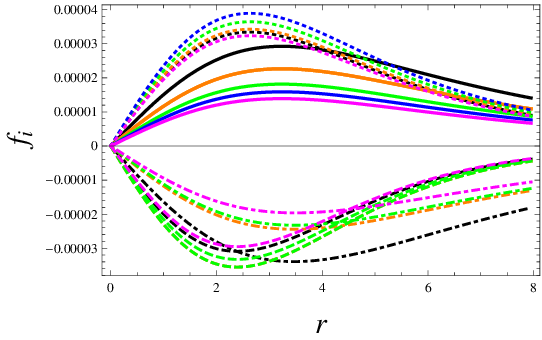,width=0.45\linewidth}\epsfig{file=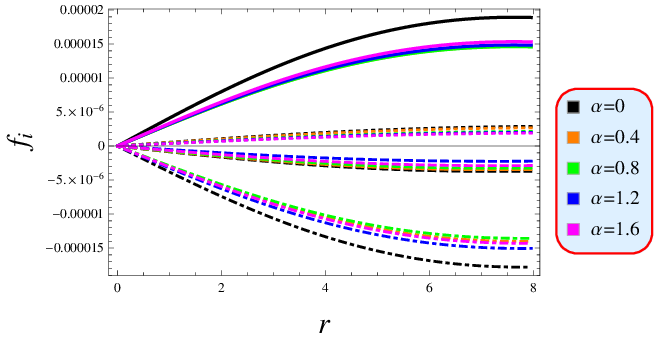,width=0.55\linewidth}
\epsfig{file=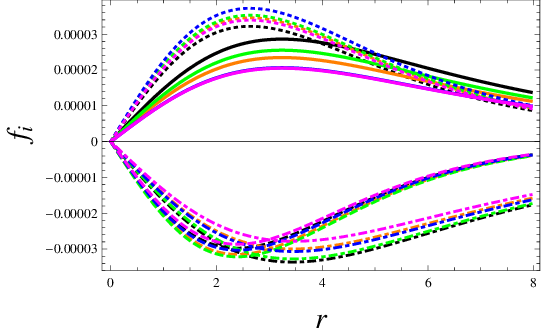,width=0.45\linewidth}\epsfig{file=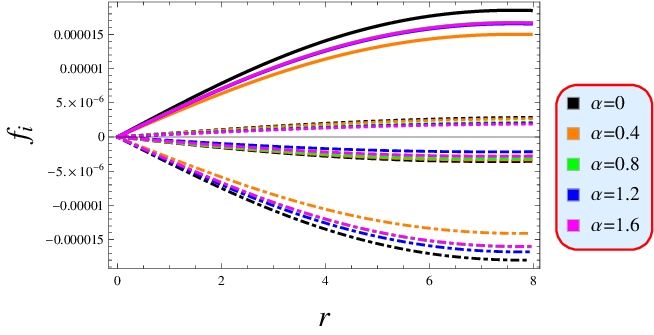,width=0.55\linewidth}
\caption{Equilibrium for solutions 1 and 2 corresponding to left and
right plots, respectively. Here, thick, dashed, dotted, and
dot-dashed lines denote $f_g$, $f_h$, $f_d$, and $f_m$,
correspondingly.}
\end{figure}

\subsubsection{Causality and Thermodynamic Behavior}

Information cannot travel faster than light, as dictated by
causality. In compact stars, the sound speed must remain below the
speed of light to preserve causality, as indicated by the theory
\cite{91}. This ensures the preservation of both equilibrium and
causality within the stellar structure. This condition is formulated
as
$$0 \leq v_s^2=\frac{dp}{d\rho} \leq 1.$$

One important indicator for assessing the system's stability is the
adiabatic index. The adiabatic index is defined as the ratio of the
relative change in pressure to the relative change in density for a
material undergoing an adiabatic process, i.e., a process that
involves no heat transfer. A value greater than $\frac{4}{3}$ for
this factor is a necessary condition for the stability of a compact
star \cite{93}. An adiabatic index below $\frac{4}{3}$ leads to
instability, threatening the star with gravitational collapse. The
mathematical representation of this component is
$$\Gamma =\bigg(1+\frac{\rho}{p}\bigg)v_s^2.$$
The fulfillment of the required conditions by both solutions is
shown in Figure \textbf{9}, which classifies them as stable
configurations.
\begin{figure}[h!]\center
\epsfig{file=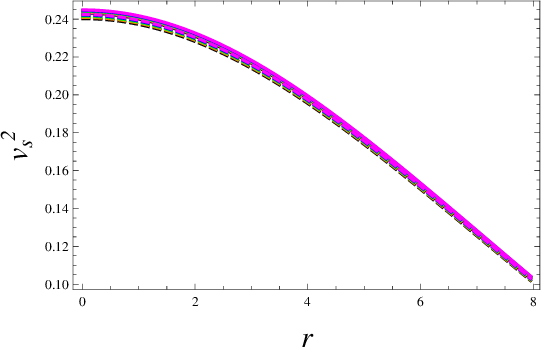,width=0.45\linewidth}\epsfig{file=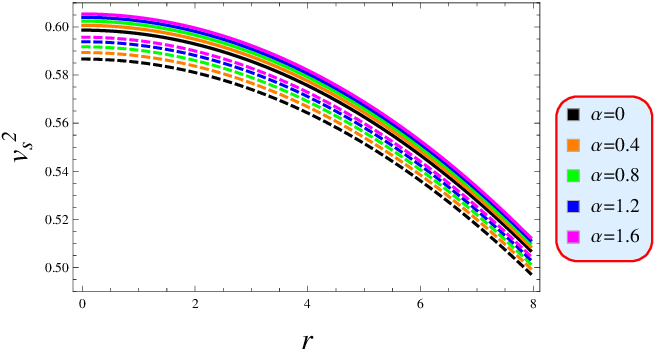,width=0.55\linewidth}
\epsfig{file=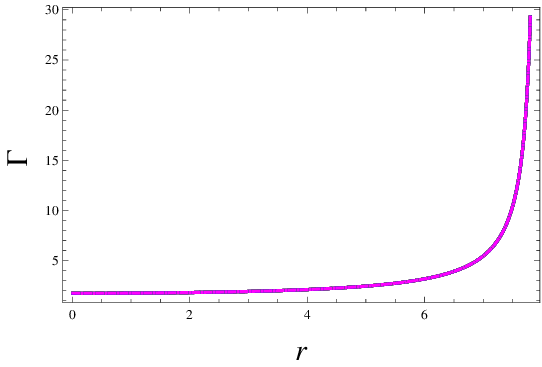,width=0.45\linewidth}\epsfig{file=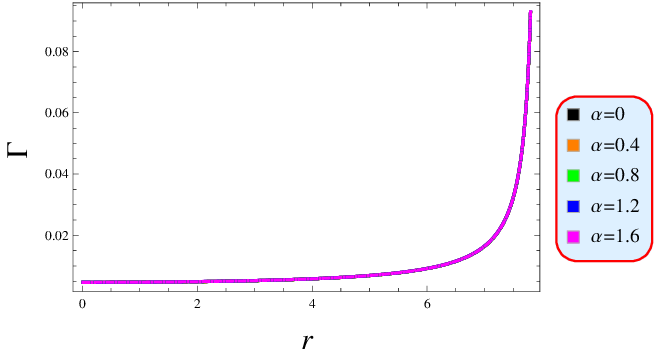,width=0.55\linewidth}
\caption{Stability for solutions 1 and 2 corresponding to left and
right plots, respectively under $\beta=1.3$ (solid) and $\beta=1.9$
(dashed).}
\end{figure}

\subsubsection{Harrison-Zeldovich-Novikov Condition}

Zeldovich's prerequisite \cite{aeb} requires the ratio of central
pressure to energy density ($p_c/ \rho_c$) to be less than one at
all points within a star. According to the
Harrison-Zeldovich-Novikov stability condition, a stable
configuration of a self-gravitating compact star must exhibit a
positive gradient of total mass with respect to its central density
throughout the entire range of the possible central density. This
stability condition is mathematically defined by the inequality
$\frac{dM}{d\rho_c} > 0$ \cite{aeb,aea}. As shown in Figure
\textbf{10}, the derivative $\frac{dM}{d\rho_c}$ exhibits a
monotonic decrease with rising central density, yet maintains
positive values throughout the entire stellar configuration. The
analysis further confirms that the ratio of the central parameters
remains below unity. Consequently, the proposed models are both
physically viable and stable.
\begin{figure}[h!]\center
\epsfig{file=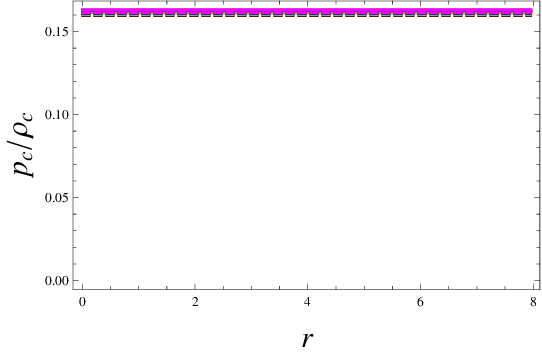,width=0.45\linewidth}\epsfig{file=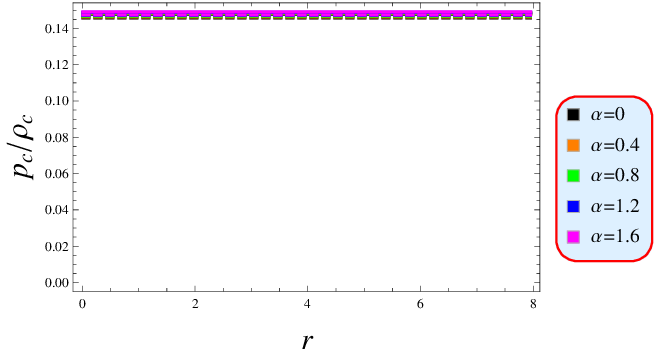,width=0.55\linewidth}
\epsfig{file=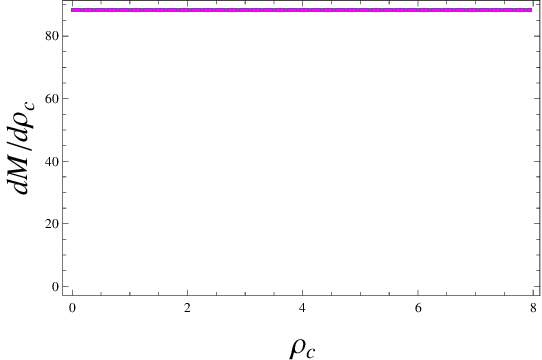,width=0.45\linewidth}\epsfig{file=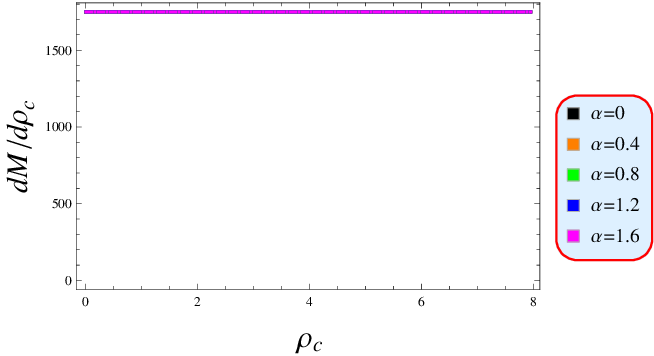,width=0.55\linewidth}
\caption{Zeldovich condition for solutions 1 and 2 corresponding to
left and right plots, respectively under $\beta=1.3$ (solid) and
$\beta=1.9$ (dashed).}
\end{figure}

\section{Conclusions}

In the framework of $f(R,L_m,T)$ gravity, this study shows two
separate solutions to the field equations for dark energy star form.
We start our investigation by deriving the extended field equations
valid for a static, spherically symmetric spacetime. From the
analysis, three non-vanishing terms emerge within the equations of
motion, corresponding to a system of seven unknowns - namely, two
metric functions alongside five matter components. The model uses
two equations of state to address this. A free parameter is
introduced in the second relation to show how the energy densities
of dark energy and ordinary matter are connected, whereas the first
relation accounts for the dark energy component itself. The number
of unknowns remained higher than the number of independent equations
in spite of these presumptions. Through an analysis of the
Kohler-Chao-Tikekar and Durgapal-Fluoria spacetime frameworks, two
distinct solutions for dark energy stars were subsequently derived.
The Schwarzschild metric was used to define the exterior geometry,
and the junction conditions at the boundary $r=R$ were then used to
find the constants in the metric potentials. These constants and the
model parameters $\alpha$ and $\beta$ were then obtained by applying
the null pressure constraint to the stellar surface. We have used
the astrophysical object SAX J 1808.4-3658 to illustrate our
findings graphically. Graphical analysis was used to further
investigate the behavior of the generated solutions at various
$\alpha$ values. The following section provides an overview of the
investigation's key conclusions.
\begin{itemize}
\item The metric potentials for both spacetimes
are clearly defined and exhibit the required positive, monotonically
increasing behavior outward from the center (Figure \textbf{1}). The
fluid parameter profiles are in line with the anticipated behavior
for a compact star object (Figure \textbf{2}). The dark energy
density declines monotonically (Figure \textbf{3}). The radial
pressure remains negative with a magnitude that increases outward,
while the tangential pressure is positive and decreasing. The
gradients of the pressure and density were also analyzed (Figure
\textbf{4}).

\item The mass function and compactness factor increase monotonically with
the radial coordinate, attaining their maximum values at the stellar
surface. In contrast, the gravitational redshift decreases toward
the center and increases outward (Figure \textbf{5}).

\item The physical feasibility of the suggested
stellar models is confirmed by the fact that both solutions meet all
energy requirements. The behavior of the equation of state parameter
is physically acceptable throughout the stellar interior (Figure
\textbf{6}).

\item The mass-radius diagram is plotted,
through which we predict the radii of three different pulsars and
find their numerical values in agreement with current observational
data (Figure \textbf{7}).

\item The star models are in hydrostatic equilibrium by showing that
the net force is zero for both configurations (Figure \textbf{8}).
Stability is maintained for both models across all parametric
choices and under each of the three stability criteria (Figures
\textbf{9} and \textbf{10}).

\item A crucial aspect of our analysis involves the systematic selection
of the free coupling parameters $\alpha$ and $\beta$ appearing in
the linear model \eqref{3c}. These parameters are not chosen
arbitrarily; rather, their admissible ranges are rigorously
constrained by the simultaneous satisfaction of a comprehensive set
of physical acceptability and stability criteria that any realistic
compact stellar configuration must obey. Specifically, we impose the
positivity and strict monotonic radial decline of the effective
energy density and pressure components, the fulfillment of all
classical energy conditions throughout the interior, the
preservation of causality for the squared sound speed, and the
dynamical stability conditions quantified by the adiabatic index,
the Zeldovich criterion at the center, and the
Harrison-Zeldovich-Novikov requirement. By applying these concurrent
constraints to our exact analytical solutions for the specific
observed parameters of the compact star SAX J1808.4-3658, we perform
a systematic numerical scan of the entire $(\alpha,\beta)$-parameter
space. The intersection of all physically permissible subspaces
unequivocally yields the allowed ranges $0 \leq \alpha \leq 1.6$ and
$1.3 \leq \beta \leq 1.9$. Within this rigorously determined viable
window, the discrete values of $\alpha$ are deliberately selected to
furnish a uniformly spaced and comprehensive survey of the
parametric dependence, where the limiting case $\alpha=0$ serves as
a crucial baseline to isolate the exclusive effects of the
trace-coupling parameter $\beta$. Concurrently, the two specific
values of $\beta$ are chosen because they precisely demarcate the
lower and upper bounds of the admissible $\beta$-domain,
respectively. This strategic selection ensures that our graphical
analysis and subsequent conclusions rigorously probe the entire
physically viable parameter spectrum, thereby validating the
robustness of our stellar models against extreme variations in the
modified gravity coupling strengths.
\end{itemize}

The identification of stellar objects in gravitational theory relies
on characteristic physical features and stability measurements. Both
of our models are consistent with these requirements. These theories
align with earlier GR research \cite{1a,1b}. Therefore, it is argued
that the existence of dark energy solutions in the $f(R,L_m,T)$
theory within a standard model is possible when appropriate
parameter values are chosen. The model replicates the usual GR
results for $\alpha = \beta = 0$.
\\\\
\textbf{Data Availability Statement:} No data is used/described in
this study.


\begin{thebibliography}{000}
\bibitem{s1} Schwarzschild, K.: Sitz. Deut. Akad. Wiss Berlin Kl. Math. Phys.
\textbf{24}(1916)424.

\bibitem{s2} Schwarzschild, K.: Sitz. Deut. Akad. Wiss Berlin Kl. Math. Phys. \textbf{1916}(1916)189.

\bibitem{s3} Capozziello, S., Stabile, A. and Troisi, A.: Class. Quantum Grav.
\textbf{25}(2008)085004.

\bibitem{s4} Nojiri, S., Odintsov, S.D. and S\'{a}ez-G\'{o}mez, D.: Phys. Lett. B \textbf{681}(2009)74.

\bibitem{s5} De Felice, A. and Tsujikawa, S.: Living Rev. Relativ.
\textbf{13}(2010)3.

\bibitem{s6} Nojiri, S. and Odintsov, S.D.: Phys. Rep. \textbf{505}(2011)59.

\bibitem{s7} Astashenok, A.V., Capozziello, S. and Odintsov, S.D.:
J. Cosmol. Astropart. Phys. \textbf{01}(2015)001.

\bibitem{s8} Rudra, P.: Commun. Theor. Phys. \textbf{66}(2016)149.

\bibitem{s9} Mustafa, G., Hussain, I. and Shamir, M.F.: Universe
\textbf{6}(2020)48.

\bibitem{s10} Rudra, P.: Nucl. Phys. B \textbf{956}(2020)115014.

\bibitem{s11} Sadeghi, J., Shokri, M., Gashti, S.N., Pourhassan, B. and
Rudra, P.: Int. J. Mod. Phys. D \textbf{31}(2022)2250019.

\bibitem{s14} Bertolami, O., Boehmer, C.G., Harko, T. and Lobo, F.S.N.:
Phys. Rev. D \textbf{75}(2007)104016.

\bibitem{s15} Pourhassan, B. and Rudra, P.: Phys. Rev. D
\textbf{101}(2020)084057.

\bibitem{s16} Naseer, T., Sharif, M., Fatima, A. and Manzoor, S.: Chin. J. Phys.
\textbf{86}(2023)350.

\bibitem{s17} Harko, T., Lobo, F.S.N., Nojiri, S. and Odintsov, S.D.:
Phys. Rev. D \textbf{84}(2011)024020.

\bibitem{15} Singh, C.P., Kumar, P.: Eur. Phys. J. C \textbf{74}(2014)11.

\bibitem{20} Sharif, M. and Zubair, M.: Astrophys. Space Sci. \textbf{349}(2014)457.

\bibitem{17} Baffou, E.H., Kpadonou, A.V., Rodrigues, M.E., Houndjo, M.J.S., Tossa, J.:
Astrophys. Space Sci. \textbf{356}(2015)173.

\bibitem{23} Kumar, P. and Singh, C.P.: Astrophys. Space Sci. \textbf{357}(2015)120.

\bibitem{33f} Rudra, P.: Eur. Phys. J. Plus \textbf{130}(2015)66.

\bibitem{33} Moraes, P.H.R.S., Arba\~{n}il, J.D.V. and Malheiro, M.: J. Cosmol. Astropart. Phys. \textbf{06}(2016)005.

\bibitem{33a} Mustafa, G., Zubair, M., Waheed, S. and Tiecheng, X.: Eur. Phys. J. C \textbf{80}(2020)26.

\bibitem{33e} Rudra, P. and Giri, K.: Nucl. Phys. B \textbf{967}(2021)115428.

\bibitem{33g} Rudra, P.: Int. J. Mod. Phys. D \textbf{31}(2022)2250095.

\bibitem{t15} Sharif, M. and Naseer, T.: Class. Quantum Grav.
\textbf{40}(2023)035009.

\bibitem{33c} Nashed, G.G.L.: Astrophys. J. \textbf{950}(2023)129.

\bibitem{tmona4} Naseer, T. et al.: Eur. Phys. J. C \textbf{84}(2024)1187.

\bibitem{ts6} Naseer, T. and Mustafa, G.: Ann. Phys.
\textbf{473}(2025)169886.

\bibitem{tfatima1} Naseer, T., Sharif, M. and Chand, F.: Phys. Scr.
\textbf{100}(2025)015014.

\bibitem{taqsa2} Sharif, M. et al.: New Astron.
\textbf{126}(2026)102552.

\bibitem{tsl7} Naseer, T.: Int. J. Geom. Methods Mod. Phys.
\textbf{23}(2026)2650045.

\bibitem{taqsa3} Sharif, M. et al.: Nucl. Phys.
B \textbf{1029}(2026)117573.

\bibitem{22a} Haghani, Z. and Harko, T.: Eur. Phys. J.
C \textbf{81}(2021)615.

\bibitem{22b} Naseer, T. and Said, J.L.: Eur. Phys. J. C \textbf{84}(2024)808.

\bibitem{89a} Zubair, M. et al.: Fortschr. Phys. \textbf{71}(2023)2300018.

\bibitem{22c} Arora, S., Moraes, P.H.R.S. and Sahoo, P.K.: Eur. Phys. J. Plus
\textbf{139}(2024)542.

\bibitem{22d} Moraes, P.H.R.S., Agrawal, A.S. and Mishra, B.: Phys. Lett. B
\textbf{855}(2024)138818.

\bibitem{22e} Pretel, J.M.Z.: Phys. Scr. \textbf{99}(2024)085001.

\bibitem{22f} Tangphati, T. et al.: Chin. Phys. C \textbf{49}(2025)025110.

\bibitem{100h} Mota, C.E., Pretel, J.M.Z. and Flores, C.O.V.: Eur. Phys. J. C
\textbf{84}(2024)673.

\bibitem{57} Ovalle, J.: Phys. Rev. D \textbf{95}(2017)104019.

\bibitem{58} Ovalle, J., Casadio, R., da Rocha, R. and Sotomayor, A.: Eur. Phys. J. C \textbf{78}(2018)122.

\bibitem{64} Ospino, J. and N\'{u}\~{n}ez, L.A.: Eur. Phys. J. C \textbf{80}(2020)166.

\bibitem{66} Ivanov, B.V.: Eur. Phys. J. C \textbf{81}(2021)227.

\bibitem{66a} Mustafa, G. Shamir, M.F. and Tie-Cheng, X.: Phys. Rev. D \textbf{101}(2020)104013.

\bibitem{67a} Sharif, M. and Naseer, T.: Phys. Scr.
\textbf{97}(2022)055004.

\bibitem{67b} Feng, Y. et al.: Chin. J. Phys. \textbf{90}(2024)372-386.

\bibitem{68} Herrera, L. et al.: J. Math. Phys. \textbf{42}(2001)2129.

\bibitem{z1} Ade, P.A.R., Aghanin, N. and Arnaud, M.: Astron. Astrophys. \textbf{594}(2016)A13.

\bibitem{z2} Ferreira, P.G.  and  Starkmann, G.:  Science \textbf{326}(2009)812.

\bibitem{z3} Abbott, B.P. et al.: Phys. Rev. \textbf{93}(2016)122003.

\bibitem{z6} Bennett, C.L. et al.: Astrophys. J. Suppl. \textbf{583}(2003)1.

\bibitem{z7} Hinshaw, G. et al.: Astrophys. J. Suppl.
\textbf{148}(2003)135.

\bibitem{z4} Riess, A.G. et al.: Astrophys. J. \textbf{560}(2001)49.

\bibitem{z5} Perlmutter, S., Turner, M.S. and White, M.J.: Phys. Rev.  \textbf{83}(1999)670.

\bibitem{z8} Riess, A.G. et al.: Astron. J. \textbf{116}(1998)1009.

\bibitem{z9} Perlmutter, S. et al.: Astrophys. J. \textbf{517}(1999)565.

\bibitem{z10} Sakti, M.F.A.R.  and  Sulaksono, A.: Phys. Rev. D \textbf{103}(2021)084042.

\bibitem{z11} Kiselev, V.V.:  Class. Quantum Grav. \textbf{20}(2003)1187.

\bibitem{z12} Fernando, S.:  Gen. Relativ. Gravit. \textbf{44}(2012)1857.

\bibitem{z13}  Xu, Z. and Wang, J.: Phys. Rev. D \textbf{95}(2017)064015.

\bibitem{z14}  Ghosh, S.G.:  Eur. Phys. J. C \textbf{76}(2016)222.

\bibitem{z15} Sakti, M.F.A.R., Suroso, A. and Zen, F.P.: Eur. Phys. J. Plus \textbf{134}(2019)580.

\bibitem{z16} Sakti, M.F.A.R., Suroso, A. and Zen, F.P.:  Annals Phys. \textbf{413}(2020)168062.

\bibitem{z17}  Bhar, P. et al.:  Eur. Phys. J. C \textbf{76}(2016)708.

\bibitem{z18}  Chernin, A.D. et al.:  Astron. Astrophys. \textbf{507}(2009)1271.

\bibitem{22bx} Das, K.P. and Debnath, U.: Int. J. Geom. Methods Mod.
Phys. \textbf{21}(2024)2450178.

\bibitem{y1} Albalahi, A.M. et al.: Eur. Phys.
J. C \textbf{84}(2024)963.

\bibitem{y3} Siza, B., Andrade, J., Santana, D., and Naseer, T.:
Eur. Phys. J. C \textbf{84}(2024)1203.

\bibitem{73} Herrera, L., Ospino, J. and Di Prisco, A.: Phys. Rev. D \textbf{77}(2008)027502.

\bibitem{71} Roupas, Z.: Astrophys. Space Sci. \textbf{366}(2021)9.

\bibitem{72} Deb, D., Mukhopadhyay, B. and Weber, F.: Astrophys. J. \textbf{922}(2021)149.

\bibitem{75} Das, S., Parida, B.K., Ray, S. and Pal, S.K.: Phys. Sci. Forum \textbf{2}(2021)29.

\bibitem{fah} Sarkar, S., Sarkar, N., Rudra, P., Rahaman, F. and
Ghorui, T.: Eur. Phys. J. C \textbf{83}(2023)1005.

\bibitem{tsl2} Naseer, T.: Eur. Phys. J. C \textbf{84}(2024)1256.

\bibitem{ts4} Demir, E. et al.: Chin. J. Phys. \textbf{91}(2024)299-315.

\bibitem{z74a} Qu, Z.S., Wang, T. and Feng, C.J.: Eur. Phys. J. C \textbf{83}(2023)784.

\bibitem{z74b} Zaslavskii, O.B.: Phys. Lett. B \textbf{875}(2026)140366.

\bibitem{z74} Kohler, M. and Chao, K.L.: Naturforsch. A \textbf{20}(1965)1537.

\bibitem{z73} Tikekar, R: Curr. Sci. \textbf{39}(1970)460.

\bibitem{nv1} Wang G, Feng Z, Qu Y and Sun, H.: PLoS One
\textbf{20}(2025)e0333700.

\bibitem{nv2} Xiang, Y. et al.: Mon. Not. R. Astron. Soc. \textbf{546}(2026)1-34.

\bibitem{nv3} Huo, M., Fan, Z., Qi, J., Qi, N. and Zhu, D.: J. Guid. Control Dyn.
\textbf{46}(2023)1015-1022.

\bibitem{nv4} Jin, J., Chen, W., Ouyang, A., Yu, F. and Liu, H.: IEEE Trans.
Emerg. Top. Comput. Intell. \textbf{8}(2024)364-376.

\bibitem{nv5} Lei, Z. et al.: SPE J. \textbf{30}(2025)743-761.

\bibitem{nv6} Liang, D., Hu, H. and Wu, Y.: Proc. Inst. Mech. Eng. K: J.
Multi-body Dyn. \textbf{239}(2025)234-253.

\bibitem{41ja} Darmois, G.: Les equations de la gravitation
einsteinienne (1927).

\bibitem{z75} Durgapal, M.C., Fuloria, R.S.: Gen. Relativ. Gravit. \textbf{17}(1985)671.

\bibitem{z77} Sharif, M. and Saba, S.: Chin. J. Phys.
\textbf{63}(2020)348.

\bibitem{z78} Ama-Tul-Mughani, Q., Salam, W. and Saleem, R.: Eur. Phys. J. Plus
\textbf{136}(2021)42.

\bibitem{89fa} Panotopoulos, G. and Lopes, I.: Phys. Rev. D
\textbf{96}(2017)083013.

\bibitem{42} Buchdahl, H.A.: Phys. Rev. \textbf{116}(1959)1027.

\bibitem{89g} Ivanov, B.V.: Phys. Rev. D \textbf{65}(2002)104011.

\bibitem{58aa} Chandrasekhar, S. and Milne, E.A.: Mon. Not. R. Astron. Soc.
\textbf{91}(1931)456.

\bibitem{59aa} Chandrasekhar, S.: Mon. Not. R. Astron. Soc. \textbf{95}(1935)207.

\bibitem{62aa} Howell, D.A. et al.: Nature \textbf{443}(2006)308.

\bibitem{4y} Demorest, P.B.: Nature \textbf{467}(2010)1081.

\bibitem{64aa} Antoniadis, J. et al.: Science \textbf{340}(2013)6131.

\bibitem{65aa} Scalzo, R.A. et al.: Astrophys. J. \textbf{713}(2010)1073.

\bibitem{66aa} Cromartie, H.T. et al.: Nat. Astron. \textbf{4}(2020)72.

\bibitem{z80} Tolman, R.C.: Phys. Rev. \textbf{55}(1939)364.

\bibitem{68aa} Bombaci, I.: Astron. Astrophys. \textbf{305}(1996)871.

\bibitem{79aa} Astashenok,A.V., Odintsov, S.D., Oikonomou, V.K.: Symmetry \textbf{15}(2023)1141.

\bibitem{80aa} Chamel, N., Haensel, P., Zdunik, J.L., Fantina, A.F.:
Int. J. Mod. Phys. E \textbf{22}(2013)1330018.

\bibitem{z81} Oppenheimer, J.R. and Volkoff, G.M.: Phys. Rev.
\textbf{55}(1939)374.

\bibitem{91} Abreu, H., Hern\'{a}ndez, H. and N\'{u}\~{n}ez, L.A.: Class. Quantum Grav.
\textbf{24}(2007)4631.

\bibitem{93} Heintzmann, H. and Hillebrandt, W.: Astron. Astrophys. \textbf{38}(1975)51.

\bibitem{aeb} Zeldovich, Y.B. and Novikov, I.D.: Relativistic Astrophysics Vol. 1:
Stars and Relativity (University of Chicago Press, Chicago, 1971).

\bibitem{aea} Harrison, B.K., Thorne, K.S., Wakano, M. and Wheeler, J.A.: Gravitational
Theory and Gravitational Collapse (University of Chicago Press,
Chicago, 1965).

\bibitem{1a} Das, K. and Ali, N.: Astrophys. Space Sci.
\textbf{356}(2015)57-66.

\bibitem{1b} Bhar, P.: Phys. Dark Universe
\textbf{34}(2021)100879.
\end{thebibliography}
\end{document}